\documentclass[a4paper,fleqn]{cas-dc}

\usepackage[numbers]{natbib}
\usepackage{amsmath,amssymb,amsthm}
\usepackage{algorithm}
\usepackage{algorithmic}
\usepackage{booktabs}
\usepackage{bm}
\usepackage{graphicx}
\usepackage{multirow}
\usepackage{subcaption}

\newtheorem{proposition}{Proposition}

\newdefinition{definition}{Definition}
\newdefinition{remark}{Remark}

\newcommand{\R}{\mathbb{R}}

\newcommand{\bw}{\bm{w}}
\newcommand{\IC}{\mathrm{IC}}
\newcommand{\SR}{\mathrm{SR}}

\newcommand{\tr}{\mathrm{tr}}

\DeclareMathOperator*{\argmax}{arg\,max}

\begin{document}
\let\WriteBookmarks\relax
\def\floatpagepagefraction{1}
\def\textpagefraction{.001}


\shorttitle{Det-MASO}
\shortauthors{}

\title[mode=title]{%
  Det-MASO: Diversity-Regularized Multi-Agent Sequential Optimization
  for Factor Generation%
}

\tnotemark[1]
\tnotetext[1]{%
  This work proposes a cooperative multi-agent sequential
  optimization framework with diversity regularization that addresses agent homogenization in continuous
  high-dimensional output spaces, with application to quantitative factor
  generation.%
}

\author[1]{Tiantian Wu}[orcid=0009-0002-4954-3869]
\ead{daeneryswu001@gmail.com}
\credit{Methodology, Software, Formal analysis, Investigation, Data curation, Visualization, Writing - original draft}
\author[1,2,3]{Peng Yang}[orcid=0000-0001-5333-6155]
\cormark[1]
\ead{yangp@sustech.edu.cn}
\credit{Conceptualization, Supervision, Writing - review editing}

\affiliation[1]{%
  organization={Department of Statistics and Data Science},
  addressline={Southern University of Science and Technology, 1088 Xueyuan Avenue, Nantou Street, Nanshan District, Shenzhen 518055, Guangdong Province, China},
  city={Shenzhen},
  postcode={518055},
  country={China}%
}

\affiliation[2]{%
  organization={Department of Computer Science and Engineering},
  addressline={Southern University of Science and Technology, 1088 Xueyuan Avenue, Nantou Street, Nanshan District, Shenzhen 518055, Guangdong Province, China},
  city={Shenzhen},
  postcode={518055},
  country={China}%
}

\affiliation[3]{%
  organization={Guangdong Provincial Key Laboratory of Brain-Inspired Intelligent Computation},
  addressline={Southern University of Science and Technology, 1088 Xueyuan Avenue, Nantou Street, Nanshan District, Shenzhen 518055, Guangdong Province, China},
  city={Shenzhen},
  postcode={518055},
  country={China}%
}

\cortext[1]{Corresponding author}


\begin{abstract}
Quantitative investment requires alpha factors that are both predictive and diverse. However, existing factor generation methods, including formulaic mining and neural end-to-end models, are mostly based on a single-agent design, which limits exploration of the factor space and reduces adaptability to changing markets. Multi-agent approaches can broaden exploration, but agents trained under a shared objective often collapse to similar outputs, while existing diversity techniques either scale poorly with asset-universe size or address redundancy only after training. We propose a multi-agent sequential optimization (Det-MASO) framework for generating diversified, context-adaptive alpha factors. It introduces a differentiable log-determinant penalty on the inter-agent correlation matrix directly into the joint objective, discouraging correlated agent behavior during training with computational cost that scales linearly with the number of assets. We further show that this penalty prevents homogeneous equilibria and preserves factor diversity throughout optimization, while a context-conditioned scheduler dynamically adjusts agent emphasis across market regimes and Shapley interaction index analysis is used to characterize the cooperation structure that emerges among them. Experiments show that Det-MASO outperforms baselines, improving the information coefficient by 1.1\%, the annualized return by 6.4\%, and the Sharpe ratio by 0.33.
\end{abstract}


\begin{highlights}
\item Det-MASO maintains diverse agent behaviors and prevents convergence to similar strategies.
\item Theory establishes repulsive gradients and effective-rank diversity bounds.
\item A context-conditioned scheduler adapts factor generation to regime shifts.
\item Det-MASO improves information coefficient (IC) and Sharpe over baselines.
\end{highlights}



\maketitle


\section{Introduction}
\label{sec:intro}

The discovery of predictive signals, called alpha factors, from financial market data is a central problem in quantitative investment\cite{qfr}. A factor maps a cross-section of asset features to a predictive score, and a portfolio's edge arises from the combined information of many factors rather than any single one. State-of-the-art tools for this task fall into two paradigms: formulaic factor mining methods\cite{alphagen,alphaforge,alphasage} and neural end-to-end factor models\cite{factorvae,RVRAE2024, xu2022stock}. Formulaic factor mining searches over spaces of symbolic expressions to produce interpretable formulas, whereas neural end-to-end factor models map raw stock features directly to continuous predictive scores without producing explicit formulas. However, factor generation in both paradigms is still treated as a single-agent problem, which remains a fundamental limitation.

A single policy, whether producing symbolic expressions or continuous neural scores, is optimized for one objective and thus yields similar factors, capturing only a narrow slice of the available predictive information. This design has two main limitations. First, it struggles to generate diverse factors during training. Formulaic methods construct factors as discrete mathematical expressions that are non-differentiable with respect to the generation policy, so redundant factors can only be removed after generation. Recent approaches such as AlphaForge\cite{alphaforge} and GFlowNet-based mining\cite{alphasage} add diversity penalties to the reward or loss function, but a single agent explores the factor space through one search strategy; the resulting diversity is confined, limiting diversity to variations within the same pattern family rather than spanning qualitatively different signals. The relationship between new and existing factors also remains implicit, as each generation is treated independently\cite{AlphaPROBE}, with no explicit mechanism to ensure that new factors complement rather than duplicate the current factor set. Neural end-to-end methods (FactorVAE\cite{factorvae}, autoencoder asset pricing\cite{autoencoder}, RVRAE\cite{RVRAE2024}, NeuralFactors\cite{neuralfactors}) face the same bottleneck: a single network converges to a single local optimum, so its factors remain narrow regardless of capacity. Second, a single policy cannot adapt its generative focus to changing markets. The informativeness of momentum, volatility, and volume signals shifts substantially across market cycles\cite{Ang2006,Moskowitz2012}, and a policy trained to maximize average performance is suboptimal in any particular regime. This rigidity accelerates alpha decay\cite{alphaagent,Falck2022}: factors discovered under one market environment lose predictive power as conditions change, yet a static policy lacks the mechanism to replace or reweight them. Different regimes require different factor strategies, suggesting a multi-agent design with regime-specific specialization as a natural response to non-stationarity. Recent work has introduced multi-agent architectures into alpha mining, but these systems assign agents to workflow roles such as generation, evaluation, and critique\cite{alphaagent,FactorMAD,rdagent,Alphalogics}, so all agents ultimately pursue the same search direction. No existing framework uses multiple agents to explore structurally diverse regions of factor space in parallel.

We address both limitations by formalizing factor generation as a
cooperative multi-agent sequential optimization problem with diversity regularization. Multiple agents, each responsible for a distinct region of the factor space, are trained jointly under a shared objective. This design immediately raises the central challenge of multi-agent learning, which is agent homogenization, the tendency of agents sharing an objective to converge to identical strategies, collapsing the multi-agent system to a single-agent
one\cite{christianos2021scaling,Li2021,SND}. In standard multi-agent benchmarks with discrete or low-dimensional action spaces, homogenization is addressed through behavioral-distance penalties (system neural diversity (SND), maximum mean discrepancy (MMD)), parameter decorrelation, or architectural
constraints. These remedies do not transfer cleanly to factor generation. Behavioral distances over $N$-dimensional asset return vectors incur $O(N^2)$ cost that scales prohibitively with the asset universe size $N$, which ranges from hundreds to thousands of assets in realistic investment universes, and post-hoc output filtering recreates the same structural weakness as the single-agent baseline. Nor do these methods address the continuous distributional drift of financial markets that makes static factor generation suboptimal across regimes.

We propose \textbf{Det-MASO}, a framework that solves the homogenization problem at its source. A key design decision is the adoption of neural end-to-end factor generation, where each agent is a differentiable neural network that maps raw stock features directly to continuous factor scores. Unlike existing neural factor models that use a single network, Det-MASO deploys multiple independent agents and embeds a differentiable diversity penalty, the negative log-determinant of the inter-agent output correlation matrix $-\log\det\Sigma(\theta)$, directly into the joint objective. This penalty requires the gradient $\partial\Sigma/\partial\theta$ to exist, which in turn requires agent outputs to be continuously differentiable with respect to their parameters. Discrete symbolic outputs fundamentally cannot satisfy this condition, while existing neural methods possess this capability but have never exploited it for multi-agent diversity enforcement. By operating in a continuous, differentiable output space with multiple agents, Det-MASO can actively discourage convergence to redundant strategies throughout training, not after the fact. This penalty is grounded in log barrier theory and scales linearly with the asset-universe size for fixed small $n$, removing the $O(N^2)$ scalability bottleneck that limits standard output-space behavioral-distance approaches. Det-MASO further incorporates a context-conditioned scheduler that reads compact multi-scale market state features and continuously redistributes generative capacity across agents in response to shifting market conditions, addressing structural non-stationarity without requiring explicit regime labels or hand-crafted switching rules. The number of agents is determined from data by clustering the intrinsic structure of market states, with each agent assigned to specialize in a distinct market regime.

Our main contributions are:
\begin{enumerate}
\item We introduce a multi-agent sequential optimization framework for neural factor generation. It adds a differentiable determinant-based diversity penalty to the joint objective to stop agents from becoming too similar, chooses the number of agents with silhouette-optimal K-means clustering of market states, and uses gradient masking with a context-conditioned scheduler to adapt to changing market conditions.

\item We show that the determinant penalty creates a repulsive gradient field that prevents homogeneous equilibria (Proposition~\ref{prop:repulsion}), and we derive a lower bound on the factor diversity kept during training (Proposition~\ref{prop:rank}).

\item Our experiments show that Det-MASO improves information coefficient (IC), annualized return, and Sharpe ratio over existing baselines. The results also show that it keeps higher factor diversity during training, supporting the value of diversity control and regime-adaptive scheduling.
\end{enumerate}


\section{Related Work}
\label{sec:related}

\subsection{Automated factor generation}
\label{sec:related_factor}

The automated discovery of alpha factors has developed along two paradigms. The first is formulaic factor mining, which searches over spaces of symbolic expressions. Evolutionary formula-mining methods\cite{autoalpha} search over expression trees to produce interpretable formulas. AlphaGen\cite{alphagen} is the first reward-driven search method to jointly optimize a factor set, using the combination model's IC as reward. AlphaForge\cite{alphaforge} decouples factor generation from combination: a generative-predictive neural network proposes candidate factors, while a dynamic combination model adjusts their weights over time. AlphaSAGE\cite{alphasage} replaces standard reinforcement learning with Generative Flow Networks (GFlowNets) and a relational graph convolutional encoder that captures the mathematical structure of alpha expressions, mitigating the mode collapse to which conventional policy-gradient methods are prone and yielding more diverse factor portfolios. The second paradigm is neural end-to-end factor generation, where deep networks map raw stock features directly to continuous predictive scores without producing explicit formulas. Representative work includes FactorVAE\cite{factorvae}, which uses a variational autoencoder to disentangle latent market factors, autoencoder asset pricing models\cite{autoencoder}, RVRAE\cite{RVRAE2024}, and NeuralFactors\cite{neuralfactors}. These methods benefit from end-to-end differentiability, but they remain single-agent systems: the lone network converges to a single local optimum, so its outputs occupy a narrow subspace of the predictive space regardless of model capacity. No existing neural factor model exploits differentiability for multi-agent diversity enforcement.

Det-MASO belongs to the neural paradigm but addresses its core limitations. It deploys multiple independent agents that explore different regions of the factor space in parallel, and embeds an explicit, differentiable diversity penalty directly on agent outputs. The choice of the neural paradigm is a technical necessity: the determinant-based penalty $-\log\det\Sigma(\theta)$ requires $\partial\Sigma/\partial\theta$ to exist, which demands continuously differentiable agent outputs, a condition that symbolic expressions cannot satisfy. While this sacrifices formula-level interpretability, Det-MASO partially compensates through a post-training Shapley interaction index analysis that provides game-theoretically grounded interpretability at the agent level.

\subsection{Diversity-Promoting Multi-Agent Learning}
\label{sec:related_multiagent}

Prior work has studied how to encourage specialization among multiple learners optimized under a shared objective. Relevant ideas appear in multi-agent learning, ensemble learning, and cooperative MARL, where a central challenge is preventing different agents or modules from collapsing to similar outputs. Existing approaches can be broadly grouped into three categories. Architectural methods induce heterogeneity through agent-specific latent codes or hypernetworks for heterogeneous parameters \cite{Wang2020roma}. Objective-based methods add diversity-promoting regularizers, such as Maximum Mean Discrepancy (MMD \cite{MDD}), System Neural Diversity (SND \cite{SND}), or mutual-information bonuses \cite{Li2021}. Theoretical studies further analyze when heterogeneous agents outperform homogeneous ones under different forms of objective aggregation \cite{Whenisdiversity}.

Although some related methods arise from cooperative MARL, our problem is factor generation with multiple agents producing high-dimensional continuous outputs over an asset universe. This distinction is important computationally. Standard output-space implementations of MMD require pairwise kernel evaluations across the asset dimension, yielding quadratic scaling in $N$ and becoming prohibitive for realistic investment universes. SND similarly operates in the full output space with quadratic dependence on $N$ and is often non-differentiable or difficult to optimize directly. By contrast, Det-MASO applies a determinant-based regularizer to the $n \times n$ inter-agent correlation matrix, with complexity $O(Nn^2 + n^3)$, which is linear in $N$ for small $n$ and therefore tractable at large asset-universe scale.

\subsection{Non-stationarity in Sequential Modeling}
\label{sec:related_nonstationary}

Financial markets exhibit structural non-stationarity arising from persistent distributional drift driven by changing microstructure, participant composition, and liquidity. This is fundamentally different from the optimization-induced non-stationarity that many multi-agent learning algorithms address. Context-conditioned methods\cite{rakelly2019efficient} handle task switches across identifiable settings, which does not match the continuous drift of financial markets. In financial applications, attention-based gating mechanisms have been explored to dynamically reweight signals as market conditions 
evolve\cite{weng2020portfolio}, but such approaches remain confined to single-agent architectures and do not address non-stationarity at the factor-generation level. In factor mining, AlphaForge\cite{alphaforge} partly addresses non-stationarity through dynamic weight adjustment in the combination model, but the factor generation process itself remains static; factors discovered under one market environment lose predictive power as conditions change, yet the generator lacks the mechanism to replace or reweight them, accelerating alpha decay\cite{alphaagent}.

Instead, Det-MASO's context-conditioned scheduler reads multi-scale market state features grounded in empirical findings\cite{Ang2006,Moskowitz2012,StiversSun2010} and dynamically reallocates generative capacity across agents, adapting the joint factor distribution to current conditions end-to-end without explicit regime labels or hand-crafted switching rules.

\subsection{Shapley values in machine learning}
\label{sec:related_shapley}

The Shapley value\cite{Shapley1953} and its extensions have been widely applied to model interpretability and feature attribution. The Shapley interaction index\cite{GrabischRoubens1999}, which captures higher-order synergistic and antagonistic effects between feature pairs, has been used to analyze neural-network interactions\cite{Sundararajan2020}. 

Det-MASO applies the Shapley interaction index to characterize the cooperation structure among agents in a multi-agent factor-generation framework and uses the portfolio IC as the characteristic function to derive a game-theoretically grounded measure of inter-agent complementarity.


\section{Method}
\label{sec:method}

\subsection{Problem Formulation}
\label{sec:formulation}

We formulate the task as a cooperative multi-agent sequential optimization problem with diversity regularization. The agents share a single system-level objective but are driven by an explicit repulsive regularizer to occupy distinct regions of the output space, in contrast to purely adversarial multi-agent setups (e.g.\ GANs) whose objectives directly oppose each other. Let $N$ denote the number of assets in the investment universe, $D$ denote the dimension of raw stock-level features, $n$ denote the number of agents, whose determination is described in Section~\ref{sec:agent_number}, and $t \in \{1, \ldots, T\}$ denote the trading-day index. The prediction target is the $\tau$-step-ahead cross-sectional forward return, defined as $r_{x,t+\tau} = \mathrm{Close}_{x,t+\tau}/\mathrm{Close}_{x,t}-1$ for asset $x$. The goal is to learn a factor function whose output scores have maximal Spearman rank correlation with $r_{t+\tau}$ across the asset cross-section:
\begin{equation}
\small
\label{eq:IC}
    \mathrm{IC}(\hat{\alpha}_t, r_{t+\tau})
    \;=\;
    \frac{\sum_{x=1}^{N} \bigl( R^{\hat\alpha}_{x,t} - \bar R^{\hat\alpha}_{t} \bigr)\bigl( R^{r}_{x,t+\tau} - \bar R^{r}_{t+\tau} \bigr)}
    {\sqrt{\sum_{x=1}^{N}\bigl( R^{\hat\alpha}_{x,t} - \bar R^{\hat\alpha}_{t} \bigr)^2
            \cdot\sum_{x=1}^{N}\bigl( R^{r}_{x,t+\tau} - \bar R^{r}_{t+\tau} \bigr)^2}},
\end{equation}
where $R^{\hat\alpha}_{x,t} = \mathrm{rank}(\hat\alpha_{x,t})$ and $R^{r}_{x,t+\tau} = \mathrm{rank}(r_{x,t+\tau})$ denote the cross-sectional ranks of the predicted factor score and the realized $\tau$-step-ahead return for asset $x$, respectively, and $\bar R^{\hat\alpha}_t = \tfrac{1}{N}\sum_{x=1}^{N} R^{\hat\alpha}_{x,t}$, $\bar R^{r}_{t+\tau} = \tfrac{1}{N}\sum_{x=1}^{N} R^{r}_{x,t+\tau}$ are their cross-sectional means. Equation~(\ref{eq:IC}) is the rank-based Pearson formulation of the Spearman correlation, which remains valid in the presence of tied ranks; the more common simplification $1 - 6\sum d^2/[N(N^2-1)]$ holds only under the no-ties assumption. $\hat{\alpha}_{\cdot,t}$ denotes the final predicted factor score, obtained by aggregating the individual agent outputs $\alpha_{1,t}, \ldots, \alpha_{n,t}$ through a context-conditioned weighting mechanism detailed in Section~\ref{sec:scheduler}. At each discrete time step $t$, every agent observes the same cross-sectional feature matrix $X_t \in \R^{N \times D}$, constructed from price and volume data available strictly prior to $t$. Each agent $i$ is parameterized by $\theta_i$, which maps the shared observation to a continuous, cross-sectionally standardized factor vector:
\begin{equation}
  \alpha_{i,t} \;=\; f_{\theta_i}(X_t) \;\in\; \R^N,
\label{eq:policy}
\end{equation}
where $f_{\theta_i} : \R^{N \times D} \to \R^N$, $i = 1,\ldots,n,$ and $f_{\theta_i}$ denotes the agent network parameterized by $\theta_i$, with $X_t$ as its input.
Each $\alpha_{i,t}$ is normalized to zero mean and unit 
variance across assets at every step. The independence of $\theta_i$ across agents is a deliberate design choice which prevents parameter sharing from enforcing structural similarity between agents, and serves as a precondition for the diversity mechanism developed below.

The combined multi-agent system produces a factor matrix $F_t(\theta) = [\alpha_{1,t},\ldots,\alpha_{n,t}] \in \R^{N \times n}$. To quantify the degree of redundancy within this matrix, we define the inter-agent output correlation matrix as
\begin{equation}
  \Sigma_t(\theta) \;=\; \tilde{F}_t(\theta)^\top \tilde{F}_t(\theta) \,/\, N
  \;\in\; \R^{n \times n}.
\label{eq:Sigma}
\end{equation}
where $\tilde{F}_t(\theta) = [\tilde\alpha_{1,t},\ldots,\tilde\alpha_{n,t}] \in \R^{N\times n}$ and each column $\tilde\alpha_{i,t} = \mathrm{Std}\!\left(\mathrm{SoftRank}_{\kappa}(\alpha_{i,t})\right)$ is the cross-sectionally standardized soft rank of agent $i$'s factor output. Here $\mathrm{SoftRank}_{\kappa}(\cdot)$ is the differentiable rank operator of \citet{Blondel2020} with temperature $\kappa$, and $\mathrm{Std}(\cdot)$ denotes cross-sectional $z$-score normalization. The diagonal entries of $\Sigma_t(\theta)$ are identically 1 by construction, while the off-diagonal entry $\Sigma_{t,ij} = \tilde\alpha_{i,t}^\top \tilde\alpha_{j,t} / N$ is the empirical \emph{rank} correlation between agents $i$ and $j$ at time $t$. Defining $\Sigma_t$ in soft-rank space rather than raw-value space ensures that the diversity penalty is measured under the same rank-based metric as the prediction target $\IC_i$ (Eq.~\ref{eq:IC}), eliminating the metric mismatch that would otherwise arise from regularizing Pearson similarity while optimizing a Spearman objective. The same soft-rank vectors $\tilde\alpha_{i,t}$ are cached and reused when evaluating $\IC_i$ during training, so the additional cost is negligible. The determinant $\det(\Sigma_t(\theta))$ serves as a natural scalar summary of collective diversity. $\det(\Sigma_t) = 1$ when all factor vectors are mutually orthogonal, and $\det(\Sigma_t) \to 0$ whenever any two agents generate linearly dependent outputs.

This formulation exposes two challenges. The first is agent homogenization. Under a shared objective, gradient ascent pushes all $\theta_i$ toward the same local optimum, driving $\det(\Sigma_t(\theta)) \to 0$ and collapsing the system to a single agent. Existing remedies either assume discrete or low-dimensional actions (SND\cite{SND}, MMD\cite{MDD}), incur $O(N^2)$ cost, or correct redundancy post-hoc (AlphaForge\cite{alphaforge}). None embeds a differentiable diversity constraint into the joint objective with cost that remains linear in $N$ for fixed small $n$. The second is structural non-stationarity. The predictive relationship between $\alpha_{i,x,t}$ and $r_{x,t+\tau}$ drifts across market regimes, and this drift persists after training. Because different regimes may favor distinct factor strategies, we choose the number of agents in a data-driven way based on the clustering structure of market-state features, as described in Section~\ref{sec:agent_number}.

\subsection{Det-MASO}
\label{sec:detmarl}

\begin{figure*}
    \centering
    \includegraphics[width=1\linewidth]{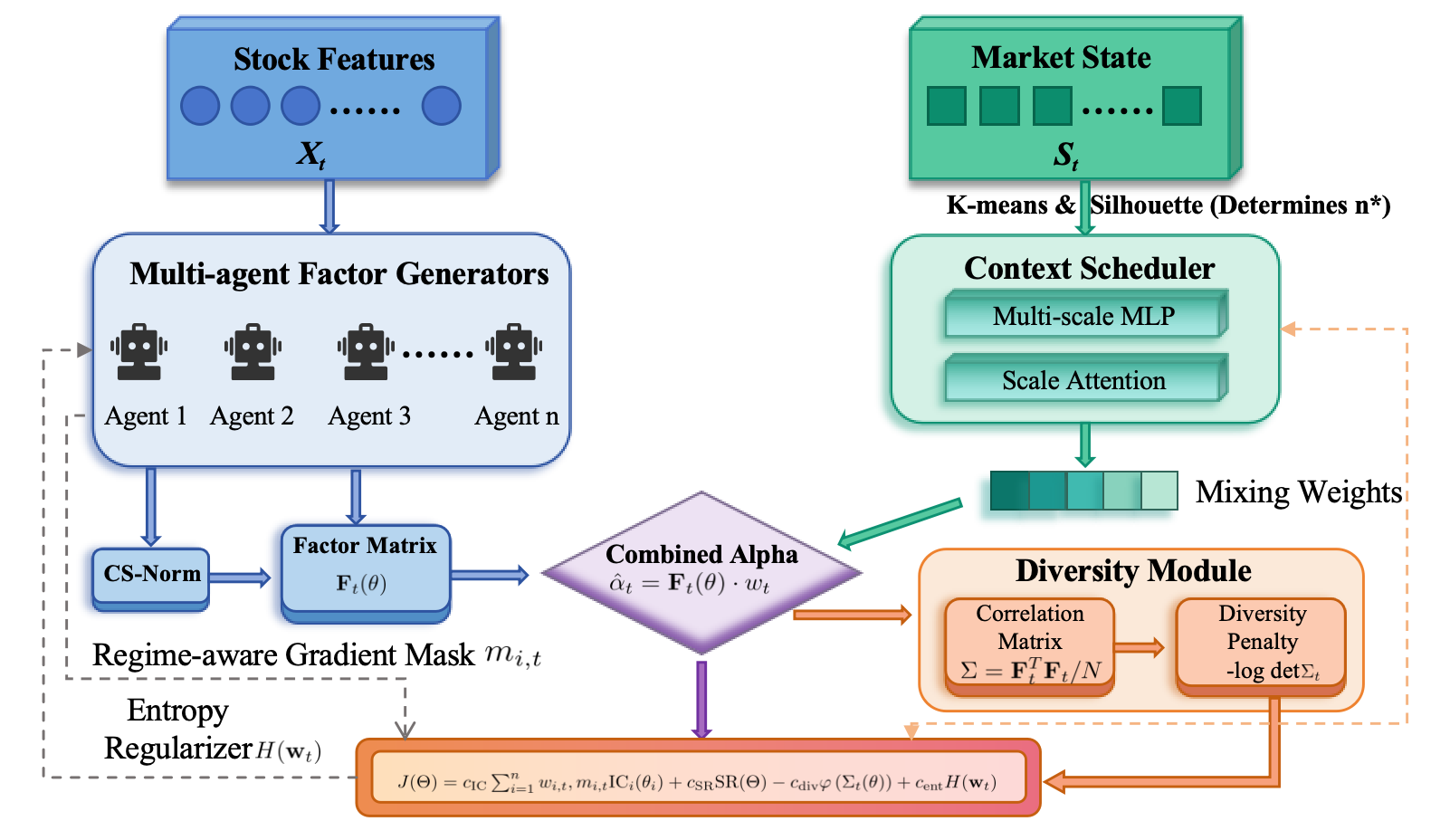}
    \caption{Each agent independently maps stock features $X_t$ to a standardized factor vector; the outputs form the factor matrix $F_t(\theta)$, from which the diversity module computes the correlation matrix $\Sigma$ and applies the penalty $-\log\det\Sigma$. A context-conditioned scheduler $f_\phi$ encodes the multi-scale market state $S_t$ via attention-fused MLPs and outputs mixing weights $w_t$ that both combine agent factors into $\hat{\alpha}_t = F_t(\theta)\cdot w_t$ and modulate per-agent gradient flow. A regime-aware gradient mask $m_i(t)$ further enforces agent–regime specialization. All components are jointly optimized under the per-step objective $J_t(\Theta)$.
}
    \label{fig:pipline}
\end{figure*}

We propose Det-MASO, a framework that addresses both challenges through four components trained end-to-end under a single objective, as illustrated in Figure~\ref{fig:pipline}. 

Unlike formulaic factor mining methods\cite{alphaagent,alphaforge,alphagen} that output discrete symbolic expressions, Det-MASO adopts neural end-to-end factor generation, where each agent is a differentiable MLP mapping raw features to continuous factor scores. This design choice is a technical prerequisite for the determinant-based diversity penalty, which requires 
$\partial\Sigma_t/\partial\theta$ to exist, a condition that symbolic outputs cannot satisfy. Unlike LLM-based multi-agent approaches (AlphaAgent\cite{alphaagent}, 
AlphaPROBE\cite{AlphaPROBE}) that rely on language-driven exploration, Det-MASO operates directly in a continuous, differentiable output space and regularizes inter-agent diversity through a differentiable determinant penalty.

\subsubsection{Determining the number of agents}
\label{sec:agent_number}
We ground the number of agents $n$ in the intrinsic clustering structure of market states, following the principle that each agent should specialize in a distinct market regime. At each trading day $t$, we construct a 12-dimensional market state vector $s_t \in \R^{12}$ from four time scales $\mathcal{W} = \{5, 20, 60, 120\}$ trading days. For each window $w \in \mathcal{W}$, three statistics are computed from the equal-weighted market portfolio returns and cross-sectional individual returns using data strictly from $[t-w, t-1]$: (i) realized volatility $\sigma_w$, (ii) time-series momentum $\mu_w$, and (iii) cross-sectional dispersion $\delta_w$ (Table~\ref{tab:features}).

\begin{table}[htbp]
\caption{Multi-scale market state features.
$r^{\mathrm{mkt}}$: equal-weighted market return.
$r_{it}$: individual asset return.
All features computed strictly from data prior to $t$, eliminating
look-ahead bias. The three statistics jointly summarize the dominant axes of market-regime variation: risk level ($\sigma_w$), directional trend ($\mu_w$), and cross-sectional return heterogeneity ($\delta_w$).}
\label{tab:features}
\begin{tabular*}{\tblwidth}{@{}L L}
\toprule
\textbf{Feature} & \textbf{Definition} \\
\midrule
Realized volatility\cite{Ang2006} & $\sigma_w = \mathrm{std}(r^{\mathrm{mkt}}_{t-w:t-1})$ \\
Time-series momentum\cite{Moskowitz2012} & $\mu_w = \sum_{k=1}^{w} r^{\mathrm{mkt}}_{t-k}$ \\
Cross-sectional dispersion\cite{StiversSun2010} & $\delta_w = \mathrm{mean}_{\tau \in [t-w, t-1]}\,\mathrm{std}_i(r_{i\tau})$ \\
\bottomrule
\end{tabular*}
\end{table}

We then apply $K$-means clustering to the set of market state vectors $\{s_t\}$ over the training period and select the number of agents by maximizing the silhouette score over candidate cluster counts $K \in \{2,\ldots,K_{\max}\}$:
\begin{equation}
  n^* = \argmax_{2 \leq K \leq K_{\max}} \;
  \frac{1}{|\mathcal{T}_{\mathrm{train}}|}
  \sum_{t \in \mathcal{T}_{\mathrm{train}}}
  \frac{b_K(s_t) - a_K(s_t)}{\max\bigl(a_K(s_t),\, b_K(s_t)\bigr)},
\label{eq:silhouette}
\end{equation}
where, under a $K$-means partition with $K$ clusters, $a_K(s_t)$ is the mean distance from $s_t$ to all other points in its own cluster and $b_K(s_t)$ is the mean distance from $s_t$ to all points in the nearest different cluster. We choose $K$-means over soft alternatives such as Gaussian Mixture Models (GMM)~\cite{Reynolds2009GMM} or Hidden Markov Models (HMM)~\cite{Rabiner1989HMM} because the regime-aware gradient mask $m_{i,t}$ (Eq.~\ref{eq:mask}) consumes hard labels, makes no distributional assumption on the derived statistics in $s_t$, and provides deterministic assignments after the centroids are estimated on the training split. We fit $K$-means once on the training split, fix the resulting centroids throughout model training, validation, and testing, and assign each day to its nearest centroid. This establishes a one-to-one correspondence between agents and market regimes, enforced during training through a regime-aware gradient mask (Section~\ref{sec:training}) that ensures each agent receives full gradient updates primarily on trading days belonging to its assigned regime.
 
The silhouette score is a parameter-free internal validity index, so this procedure requires no thresholds to tune. The agent count is derived directly from the structure of market states, ensuring that the number of agents matches the number of regimes the scheduler must adapt to. After selecting the silhouette-optimal count, we set $n = n^*$ and use $n$ to denote the realized number of agents throughout the remainder of the method and experiments. For notational simplicity in the regime-aware mask (Eq.~\ref{eq:mask}), we re-index the $K$-means cluster labels to $\{1, \ldots, n\}$ and adopt the identity mapping between agents and regimes, so that agent $i$ is assigned to regime $i$.

\subsubsection{Agent architecture and feature engineering}
\label{sec:architecture}

Each agent $i$ is a 3-hidden-layer MLP that maps $D$ input features to a scalar output per asset. Weights are shared across the $N$ stocks but completely independent across agents. The scalar outputs are collected across all $N$ assets and standardized to zero mean and unit variance cross-sectionally. Architecture details are given in Section~\ref{sec:experiments}.

The input feature matrix $X_t \in \R^{N \times D}$ has $D = 34$ features, all cross-sectionally $z$-score normalized at each date using only data from $t-1$ and before. The features span nine categories designed to capture distinct aspects of cross-sectional return predictability, summarized in Table~\ref{tab:stock_features}.

\begin{table}[htbp]
    \centering
    \renewcommand{\arraystretch}{1.2}
    \caption{Stock-level features ($D=34$). All features are cross-sectionally $z$-score standardized at each date. The nine categories are chosen to jointly span complementary channels of cross-sectional return predictability documented in the empirical asset-pricing literature, such as momentum, risk, liquidity, technical trend, intraday structure, and higher moments, rather than to optimize any single anomaly.}
    \label{tab:stock_features}
    \small
    \begin{tabular}{@{} p{2.5cm} p{5.2cm} @{}}
        \toprule
        \textbf{Category} & \textbf{Features} \\
        \midrule
        Return momentum & 1/5/10/20/60/120-day cumulative return \\
        Realized volatility & 5/10/20/60-day return std.\ deviation \\
        Volume momentum & 5/10/20/60-day relative turnover \\
        Price--volume correlation & 5/10/20-day Spearman correlation \\
        Relative Strength Index (RSI) proxy & 5/10/20-day up-day fraction \\
        Moving average (MA) deviation & 5/20/60-day MA deviation \\
        Moving Average Convergence Divergence (MACD) type & $\mu_{20}-\mu_{60}$ (short minus long cumulative return) \\
        Intraday structure & Overnight return, intraday return, close position, high--low spread $\times 2$ \\
        Statistical moments & 60-day skewness, lag-1 autocorrelation (60d), momentum Sharpe $\mu_{60}/\sigma_{60}$, 20-day high--low range, 20-day log volume \\
        \bottomrule
    \end{tabular}
\end{table}
The $N$ stocks' feature vectors are processed in parallel by each agent, and the outputs are concatenated into the joint factor matrix:
\begin{equation}
  F_t(\theta) = [\alpha_{1,t}, \ldots, \alpha_{n,t}] \in \R^{N \times n}.
\label{eq:factor_matrix}
\end{equation}

\subsubsection{Multi-scale market state representation}
\label{sec:features}

The scheduler reads a compact representation of the current market environment to determine how generative capacity should be allocated across agents. We construct the market state vector $s_t \in \R^{12}$ from four time scales $\mathcal{W} = \{5, 20, 60, 120\}$ trading days, corresponding to the canonical weekly, monthly, quarterly, and semi-annual decision horizons. For each window $w \in \mathcal{W}$, three statistics are computed whose relationship to cross-sectional factor efficacy is established in the empirical literature (Table~\ref{tab:features}), yielding a $4 \times 3 = 12$ dimensional state vector.

Features from the four scales are fused via a lightweight attention mechanism\cite{hu2024overview}. Let $s_t^w \in \R^3$ denote the sub-vector of $s_t$ corresponding to window $w$, and let $h_w = \mathrm{MLP}_w(s_t^w) \in \R^{16}$ be its encoding through a scale-specific two-layer MLP. The attention weights and fused representation are:
\begin{equation}
\small
  \bm{a}_t = \mathrm{softmax}\!\left(W_a \cdot
    \mathrm{concat}(h_{w_1}, h_{w_2}, h_{w_3}, h_{w_4})\right),
\label{eq:attn_weights}
\end{equation}
\begin{equation}
  h_t = \sum_{k=1}^{4} a_{k,t} \cdot h_{w_k} \in \R^{16},
\label{eq:fused}
\end{equation}
where $\bm{a}_t\in \R^4$ and $W_a \in \R^{4 \times 64}$ is the attention projection matrix. The attention weights $\bm{a}_t$ reveal which time scale the scheduler relies on under different market conditions, analyzed in Section~\ref{sec:nonstationarity}.

\subsubsection{Context-conditioned agent scheduler}
\label{sec:scheduler}

The fused representation $h_t$ is passed to the agent weight network $f_\phi$, which outputs the mixing weights over the $n$ agents:
\begin{equation}
  \bw_t = \mathrm{softmax}\!\left(
    W_2 \cdot \mathrm{ReLU}(W_1 h_t + b_1) + b_2
  \right) \in \Delta^{n-1},
\label{eq:scheduler}
\end{equation}
where $W_1 \in \R^{32 \times 16}$, $W_2 \in \R^{n \times 32}$, and $\Delta^{n-1}$ denotes the $(n-1)$-simplex, guaranteeing $\sum_{i=1}^{n} w_{i,t} = 1$. The scheduler's output layer is initialized to zero so that $\bw_0$ is uniform across agents at the start of training.

The weight $w_{i,t}$ plays two roles. First, it combines the agent outputs into the final predicted factor:
\begin{equation}
  \hat{\alpha}_t = F_t(\theta) \cdot \bw_t = \sum_{i=1}^{n} w_{i,t} \cdot \alpha_{i,t} \in \R^N.
\label{eq:prediction}
\end{equation}
Second, it modulates the gradient flow to each agent:
\begin{equation}
  \frac{\partial J_t}{\partial \theta_i}
  \;=\;
  c_{\text{IC}}\,w_{i,t}\,m_{i,t}\,
  \frac{\partial\,\IC_i(\theta_i)}{\partial \theta_i}
  \;+\;
  c_{\text{SR}}\,\frac{\partial\,\SR(\Theta)}{\partial \theta_i}
  \;-\;
  c_{\text{div}}\,\frac{\partial\,\varphi(\Sigma_t(\theta))}{\partial \theta_i},
\label{eq:grad_scaling}
\end{equation}
where the entropy term does not appear because $H(\bw_t)$ depends only on the scheduler parameters $\phi$. The first term makes explicit that the IC-driven gradient received by agent $i$ is scaled by both the scheduler weight $w_{i,t}$ and the regime-aware mask $m_{i,t}$, while the Sharpe term and the diversity regularizer provide additional gradients through the aggregated factor $\hat{\alpha}_t$ and the inter-agent rank-correlation structure, respectively. Thus, agents with higher $w_{i,t}$ receive stronger task-aligned updates, shifting generative emphasis toward the agent whose assigned regime matches the current market conditions.

All features in $s_t$ are computed from data in $[1, t-1]$ only. The $K$-means centroids used for regime assignment are fitted once on the training split and then held fixed throughout model training, validation, and testing, eliminating look-ahead bias.

\subsubsection{Diversity-regularized joint objective}
\label{sec:reward}

We define the joint training objective as:
\begin{equation}
\begin{split}
  J_t(\Theta) \;=\;&\;
  c_{\text{IC}} \sum_{i=1}^{n} w_{i,t}\,m_{i,t}\,\IC_i(\theta_i)
  \;+\; c_{\text{SR}}\,\SR(\Theta) \\
  &\;-\; c_{\text{div}}\,\varphi\!\left(\Sigma_t(\theta)\right)
  \;+\; c_{\text{ent}}\,H(\bw_t),
\end{split}
\label{eq:reward}
\end{equation}
where the cross-sectional terms $\IC_i$, $\Sigma_t$, $H(\bw_t)$ are evaluated at time $t$, and $\SR(\Theta)$ is the rolling-window Sharpe ratio over $\mathcal{T}_t = \{\max(1, t-T_{\mathrm{train}}+1), \ldots, t\}$ (made explicit in Eq.~\ref{eq:sharpe}). $c_{\text{IC}}, c_{\text{SR}}, c_{\text{div}}, c_{\text{ent}} > 0$ are scalar hyperparameters whose values are given in Section~\ref{sec:experiments}. $\IC_i(\theta_i)$ is the rank correlation between agent $i$'s factor and $\tau$-step-ahead returns, formally $\IC_i(\theta_i) = \mathrm{Spearman}(\alpha_{i,t},\, r_{t+\tau})$. $\SR(\Theta)$ is the Sharpe ratio of the portfolio constructed 
from $\hat{\alpha}_t$:
\begin{equation}
  \SR(\Theta)
  \;=\;
  \frac{\frac{1}{|\mathcal{T}_t|}\sum_{s \in \mathcal{T}_t}
        \sum_{x=1}^{N} \hat{\alpha}_{x,s}\, r_{x,s+\tau}}
       {\mathrm{std}_{s \in \mathcal{T}_t}\!\!\left(
        \sum_{x=1}^{N} \hat{\alpha}_{x,s}\, r_{x,s+\tau}\right)},
\label{eq:sharpe}
\end{equation}
where $\mathcal{T}_t = \{\max(1, t-T_{\mathrm{train}}+1), \ldots, t\}$ denotes the rolling training window ending at date $t$, and $\hat{\alpha}_{x,s}$ serves as the portfolio weight for asset $x$ at time $s$.
$\bw_t = [w_{1,t},\ldots,w_{n,t}]$ are time-varying agent mixing weights generated by the scheduler (Section~\ref{sec:scheduler}), and $m_{i,t}$ is the 
regime-aware gradient mask. The diversity penalty is 
$\varphi(\Sigma_t(\theta)) = -\log\det(\Sigma_t(\theta))$, and the scheduler entropy regularizer is:
\begin{equation}
  H(\bw_t) = -\sum_{i=1}^{n} w_{i,t} \log w_{i,t},
\label{eq:entropy}
\end{equation}
which prevents the scheduler weights from collapsing to a single agent.

The regime-aware gradient mask enforces agent--regime specialization:
\begin{equation}
  m_{i,t}
  \;=\;
  \mathbf{1}\!\bigl[\mathrm{regime}(t) = i\bigr]
  \;+\;
  \lambda_{\mathrm{mask}}\,\mathbf{1}\!\bigl[\mathrm{regime}(t) \neq i\bigr],
\label{eq:mask}
\end{equation}
where $\mathrm{regime}(t) \in \{1, \ldots, n\}$ is the cluster assignment of day $t$'s market state, $\mathbf{1}[\cdot]$ is the indicator function, and 
$\lambda_{\mathrm{mask}} \in [0, 1]$ controls gradient attenuation for non-assigned regimes. At one extreme $\lambda_{\mathrm{mask}} = 0$ gives hard specialization; at 
the other $\lambda_{\mathrm{mask}} = 1$ recovers the unmasked baseline. Intermediate values let each agent develop expertise in its assigned regime while retaining a weak gradient signal from others to prevent catastrophic forgetting.

The choice of $-\log\det(\Sigma_t)$ is the canonical log-barrier function for the positive-definite cone \citep{Boyd2004} and satisfies three properties:
$-\log\det(\Sigma_t) \to +\infty$ as $\det(\Sigma_t) \to 0$, inducing a repulsive gradient that prevents homogeneous configurations; strict convexity in $\Sigma_t$ on the positive-definite cone, giving the regularizer a well-behaved repulsive structure even though the overall optimization in $\theta$ remains non-convex; and a componentwise gradient
\[
  \frac{\partial}{\partial \theta_k}\bigl[-\log\det(\Sigma_t)\bigr]
  \;=\;
  -\,\mathrm{tr}\!\left(
    \Sigma_t^{-1}\frac{\partial \Sigma_t}{\partial \theta_k}
  \right),
\]
which depends on the inverse of only the $n \times n$ matrix $\Sigma_t$, keeping the regularization cost at $O(Nn^2 + n^3)$ per step, which is linear in $N$ for fixed small $n$. A comparison with alternative diversity measures is provided in Table~\ref{tab:complexity}. For the theoretical analysis below, we fix a time step $t$ and write $J(\Theta) := J_t(\Theta)$ to simplify notation.

\begin{table}[h]
\caption{Complexity comparison of diversity regularization measures. $N$: number of assets; $n$: number of agents. $N \gg n$ in all experiments.}
\label{tab:complexity}
\begin{tabular*}{\tblwidth}{@{}L L c c@{}}
\toprule
\textbf{Measure} & \textbf{cost} & \textbf{\begin{tabular}[c]{@{}c@{}}Differen-\\tiable\end{tabular}} & \textbf{\begin{tabular}[c]{@{}c@{}}Linear\\in $N$\\(fixed $n$)\end{tabular}} \\
\midrule
MMD   & $O(N^2)$       & Yes            & No \\
SND              & $O(N^2)$       & No             & No \\
$-\log\det\Sigma$ (ours) & $O(Nn^2 + n^3)$ & Yes      & Yes \\
\bottomrule
\end{tabular*}
\end{table}

The following results establish the theoretical soundness of the regularizer.

\begin{proposition}[Repulsion from homogeneous equilibria]
\label{prop:repulsion}
Let $\bar{\theta}$ be a homogeneous equilibrium, i.e.\
$\theta_i = \theta_j$ for all $i \neq j$. Assume:
\begin{itemize}
  \item[\textup{(A1)}] The map $\theta \mapsto \Sigma(\theta)$ is
  continuous at $\bar{\theta}$.
\end{itemize}
Then $J(\bar{\theta}) = -\infty$, and no gradient-ascent trajectory
$\{\theta^{(k)}\}$ with $\det(\Sigma(\theta^{(0)})) > 0$ and
monotonically non-decreasing $J$ (guaranteed under standard step-size
control) can converge to $\bar{\theta}$.
\end{proposition}

\begin{proof}
At $\bar{\theta}$, all agents produce identical standardized factors,
so every pairwise correlation equals one and
$\Sigma(\bar{\theta}) = \mathbf{1}\mathbf{1}^\top$ has rank $1$ (for
$n \geq 2$), giving $\det(\Sigma(\bar{\theta})) = 0$ and hence
$\varphi(\Sigma(\bar{\theta})) = -\log 0 = +\infty$. The remaining
terms in $J$---the IC sum $\sum_i w_{i,t} m_{i,t}\,\mathrm{IC}_i$,
the Sharpe term $\mathrm{SR}(\theta)$, and the entropy
$H(\mathbf{w}_t)$---are all bounded. With $c_{\mathrm{div}} > 0$ and $\text{B} = c_{\text{IC}} \sum_{i=1}^{n} w_{i,t}\,m_{i,t}\,\IC_i(\theta_i)+ c_{\text{SR}}\,\SR(\Theta)+c_{\text{ent}}\,H(\bw_t)$,
\begin{equation}
  J(\bar{\theta}) \;=\; \text{B}
  \;-\; c_{\mathrm{div}} \cdot (+\infty) \;=\; -\infty.
\end{equation}

Suppose for contradiction that some gradient-ascent trajectory
$\{\theta^{(k)}\}$ with $\det(\Sigma(\theta^{(0)})) > 0$ satisfies
$\theta^{(k)} \to \bar{\theta}$. By~\textup{(A1)},
$\Sigma(\theta^{(k)}) \to \mathbf{1}\mathbf{1}^\top$.
Since $\Sigma \mapsto -\log\det\Sigma$ is continuous on the open
positive-definite cone and blows up at its boundary where
$\det\Sigma = 0$, we have
$\varphi(\Sigma(\theta^{(k)})) \to +\infty$, and therefore
$J(\theta^{(k)}) \to -\infty$.
However, monotonicity of $J$ along the trajectory implies
$J(\theta^{(k)}) \geq J(\theta^{(0)}) > -\infty$ for all $k$,
contradicting $J(\theta^{(k)}) \to -\infty$.
Hence $\bar{\theta}$ is not a limit point of any such trajectory.
\end{proof}

\begin{remark}[Divergence rate of the gradient]
\label{rem:rate}
The function-value argument above already rules out convergence to
$\bar{\theta}$, but the \emph{repulsive} nature of the regularizer can
be quantified. Writing $\eta_i = \theta_i - \bar{\theta}$, a
second-order expansion under a non-degenerate parameterization yields
\begin{equation}
  1 - \rho_{ij}(\theta)
  \;=\; \tfrac{\kappa}{2}\,
  (\eta_i - \eta_j)^{\!\top} M\,(\eta_i - \eta_j)
  \;+\; O(\|\eta\|^3),
\end{equation}
for some $M \succ 0$ and $\kappa > 0$.
For $n = 2$, letting $u = \eta_1 - \eta_2$, this gives
$1 - \rho \sim \|u\|_M^2$ and
\begin{equation}
  \frac{\partial}{\partial \eta_1}
  \!\left[-\log(1-\rho^2)\right]
  \;=\; \frac{2\rho}{1-\rho^2}\cdot\frac{\partial \rho}{\partial \eta_1}
  \;\sim\; \frac{1}{\|u\|_M},
\end{equation}
so $\|\nabla_{\theta_i}\varphi(\Sigma)\| = \Omega(1/\|\eta\|)$ as
$\|\eta\| \to 0$. An analogous bound holds for general $n$.
\end{remark}

\begin{remark}[Scope of the guarantee]
\label{rem:scope}
Proposition~\ref{prop:repulsion} establishes that gradient ascent
cannot reach a homogeneous equilibrium. It does not assert
convergence to a globally optimal diverse equilibrium, as the joint
optimization landscape is non-convex; sufficiency is validated
empirically in Section~\ref{sec:diversity_analysis}, where we track
$\det(\Sigma(\theta))$ throughout training and compare against both
non-regularized baselines and behavioral-distance alternatives
(SND, MMD), demonstrating that the determinant penalty maintains
diversity at lower computational cost across both datasets.
\end{remark}

\begin{remark}[Assumptions]
\label{rem:assumptions}
Assumption~\textup{(A1)} is satisfied by all smooth activation
functions (GELU, ELU, Tanh) and almost surely by ReLU under
continuous input distributions, so it is mild and standard in the
neural-network optimization literature. The addition of the entropy
regularizer $H(\mathbf{w}_t)$ and the gradient mask $m_{i,t}$ does
not affect Proposition~\ref{prop:repulsion}: $H(\mathbf{w}_t)$ is
bounded, so the divergence of $\varphi$ dominates; and
$m_{i,t} \geq \lambda_{\mathrm{mask}} > 0$ for any
$\lambda_{\mathrm{mask}} > 0$, so the IC gradient is attenuated but
never zeroed.
\end{remark}

\begin{proposition}[Effective rank lower bound]
\label{prop:rank}
Suppose training converges to a parameter configuration satisfying
$$\det(\Sigma(\theta)) \geq \varepsilon > 0.$$
Then the smallest eigenvalue of $\Sigma(\theta)$ satisfies:
\begin{equation}
  \lambda_{\min}(\Sigma) \;\geq\; \frac{\varepsilon}{n^{n-1}},
\end{equation}
and consequently $\Sigma(\theta)$ has full rank $n$.
In particular, the factor matrix $F(\theta)$ spans an $n$-dimensional
subspace of the asset return space.
\end{proposition}

\begin{proof}
Since $\Sigma(\theta)$ is a correlation matrix, all diagonal entries equal 1,
so $\tr(\Sigma) = n$.
For any positive semidefinite matrix, each eigenvalue is bounded above by the
trace, giving $\lambda_i \leq n$ for all $i$.
Writing $\det(\Sigma) = \prod_{i=1}^{n} \lambda_i$ and applying the upper
bound to all factors except $\lambda_{\min}$:
\begin{equation}
  \varepsilon \;\leq\; \det(\Sigma)
  \;=\; \lambda_{\min} \cdot \prod_{i=2}^{n} \lambda_i
  \;\leq\; \lambda_{\min} \cdot n^{n-1}.
\end{equation}
Rearranging gives $\lambda_{\min} \geq \varepsilon/n^{n-1}$.
Since $\varepsilon > 0$, we have $\lambda_{\min} > 0$, so $\Sigma(\theta)$ is
positive definite and has full rank $n$.
\end{proof}

Propositions~\ref{prop:repulsion} and~\ref{prop:rank} together provide a
complete picture of what the determinant regularizer guarantees:
Proposition~\ref{prop:repulsion} establishes that gradient ascent cannot
reach any configuration where $\det(\Sigma) = 0$, while
Proposition~\ref{prop:rank} quantifies the diversity that must be maintained
at any configuration the optimizer does reach.
For our setting ($n \leq 5$, $\varepsilon$ calibrated on the validation
split), the bound $\varepsilon/n^{n-1}$ provides a theoretical guarantee of full rank;
the tightness of this bound is validated empirically by tracking
$\lambda_{\min}(\Sigma(\theta))$ directly on the training curves in
Section~\ref{sec:diversity_analysis}.

\subsubsection{End-to-end training.}
\label{sec:training}
All agent networks $\{\theta_i\}$ and the scheduler $f_\phi$ are collected
into a single parameter vector $\Theta = (\theta_1,\ldots,\theta_{n},
\phi)$ and updated jointly by a single Adam optimizer.
At update date $t$, we minimize the negative joint objective:
\begin{equation}
\begin{split}
  \mathcal{L}_t(\Theta) \;=\;&\;
  -\,c_{\text{IC}} \sum_{i=1}^{n} w_{i,t}\,m_{i,t}\,\IC_i(\theta_i)
  \;-\; c_{\text{SR}}\,\SR(\Theta) \\
  &\;+\; c_{\text{div}}\,\bigl[-\log\det(\Sigma_t(\theta))\bigr]
  \;-\; c_{\text{ent}}\,H(\bw_t).
\end{split}
\label{eq:loss}
\end{equation}

Since the Spearman rank correlation is not differentiable with respect to $\alpha_{i,t}$, we substitute a differentiable soft-rank approximation during training \citep{Blondel2020}, using a base temperature $\kappa_{\mathrm{soft}}$ scaled by the cross-section size as $\kappa_{\mathrm{scaled}} = \kappa_{\mathrm{soft}} \cdot N / 100$. Exact Spearman IC is reported at evaluation. The Sharpe term uses $\hat{\alpha}_{x,t}$ directly as the portfolio weight, so it remains differentiable throughout. When computing the gradient of $\SR(\Theta)$ at update date $t$, the portfolio weights $\hat{\alpha}_s$ for historical steps $s < t$ in the rolling window $\mathcal{T}_t$ are treated as fixed constants (stop-gradient), so updates act only on the current parameters without requiring backpropagation through time.

The learning rate follows a cosine annealing schedule:
\begin{equation}
  \mathrm{lr}(t) = \eta_{\min} + \frac{1}{2}(\eta_{\max} - \eta_{\min})\left(1 + \cos\frac{t\pi}{T_{\max}}\right),
\label{eq:cosine}
\end{equation}
with $\eta_{\max} = 10^{-3}$, $\eta_{\min} = 10^{-5}$, and $T_{\max} = |\mathrm{train\_dates}| \times \mathrm{epochs}$.
Early stopping is applied based on the 3-epoch smoothed validation IC with patience 8, restoring the best-epoch model weights.

Training proceeds in a rolling-window fashion at the parameter-update level: at each rebalance date the
model is updated on the $T_{\mathrm{train}}$ most recent observations, with
strictly no access to future data. The regime centroids remain fixed after being estimated on the training split.
The complete procedure is given in Algorithm~\ref{alg:training}.

\begin{algorithm}[htbp]
\caption{Det-MASO Training}
\label{alg:training}
\begin{algorithmic}[1]
\REQUIRE Features $\{X_t\}$, returns $\{r_{t+\tau}\}$; coefficients $c_{\text{IC}},c_{\text{SR}},c_{\text{div}},c_{\text{ent}},\lambda_{\mathrm{mask}}$; $T_{\mathrm{train}}, E_{\max}, P, K_{\max}$
\ENSURE Agent networks $\{f_i\}_{i=1}^{n}$, scheduler $f_{\phi}$, Shapley indices $\{I(i,j)\}_{i \neq j}$
\STATE $n \leftarrow \argmax_{2 \leq K \leq K_{\max}}\! \mathrm{Silhouette}(K)$ on $\{s_t\}$ \hfill\COMMENT{Eq.~\ref{eq:silhouette}; here $n = n^*$}
\STATE Initialise $\Theta = (\theta_1,\ldots,\theta_{n},\phi)$
\FOR{$e = 1, \ldots, E_{\max}$}
\FOR{$t = 1, \ldots, T_{\mathrm{train}}$}
  \STATE Compute $s_t$ from data in $[1, t-1]$
  \STATE $h_t \leftarrow \sum_k a_{k,t} \cdot h_{w_k}$ \hfill\COMMENT{Eq.~\ref{eq:attn_weights}--\ref{eq:fused}}
  \STATE $\bw_t \leftarrow f_{\phi}(h_t)$;\ $\alpha_{i,t} \leftarrow f_i(\theta_i; X_t)$
  \STATE $\hat{\alpha}_t \leftarrow \sum_i w_{i,t}\,\alpha_{i,t}$;\ compute $m_{i,t}$ \hfill\COMMENT{Eq.~\ref{eq:prediction},\,\ref{eq:mask}}
  \STATE Compute $\IC_{i,t}, \SR, \Sigma_t, H(\bw_t)$
  \STATE Update $\Theta$ via Adam on $\mathcal{L}_t(\Theta)$ \hfill\COMMENT{Eq.~\ref{eq:loss}}
\ENDFOR
\STATE Early stop on smoothed val IC (patience $P$); \textbf{break} if triggered
\ENDFOR
\STATE Compute $\{I(i,j)\}_{i \neq j}$ exactly by exhaustive evaluation of scheduler-reweighted $v(S)$ \hfill\COMMENT{Eq.~\ref{eq:shapley_v}, \ref{eq:shapley}}
\end{algorithmic}
\end{algorithm}

\noindent\textbf{Complexity.}
The per-step training cost is $O(NDn)$, dominated by the factor generation
forward pass, identical in order to a single-agent baseline of equivalent
depth.
For our setting ($N = 500$, $D = 34$, $n \leq 5$) this amounts to
approximately $8.5 \times 10^4$ multiply-accumulate operations per step.
The log-det regularization costs $O(Nn^2 + n^3)$ per step, which is linear in $N$ for fixed small $n$,
and is therefore substantially more efficient in the asset dimension than quadratic-in-$N$ output-space MMD and SND (see
Table~\ref{tab:complexity}).
The post-training Shapley computation evaluates the exact characteristic function over all $2^{n}$ subsets. With $n=5$, this requires evaluating only 32 subset combinations on the evaluation period, computing instantly on a single GPU without requiring Monte Carlo approximation.

\subsubsection{Characterizing the emergent ecosystem via the Shapley Interaction Index.}
\label{sec:shapley}
The diversity regularizer ensures that agents do not collapse to identical
outputs.
However, preventing homogenization does not reveal \textit{how} agents relate
to one another once training has converged.
To characterize the emergent cooperation structure, we apply the
\textit{Shapley Interaction Index} \citep{GrabischRoubens1999} as a
post-training analysis tool.
Define the time-averaged characteristic function $v : 2^{[n]} \to \R$ as
\begin{equation}
\label{eq:shapley_v}
\begin{split}
  v(S) &\;=\; \frac{1}{T_{\mathrm{eval}}} \sum_{t \in \mathcal{T}_{\mathrm{eval}}} \IC\!\left( \hat{\alpha}^{S}_{t},\, r_{t+\tau} \right), \\
  \hat{\alpha}^{S}_{t} &\;=\; \sum_{i \in S} \frac{w_{i,t}}{\sum_{j \in S} w_{j,t}} \cdot \alpha_{i,t}, \quad \forall\, S \subseteq [n],\ S \neq \emptyset,
\end{split}
\end{equation}
where $\mathcal{T}_{\mathrm{eval}}$ is the evaluation period with cardinality $T_{\mathrm{eval}}$, $\IC(\cdot,\cdot)$ is the Spearman rank correlation defined in Eq.~\ref{eq:IC}, and $\hat{\alpha}^{S}_{t}$ is the scheduler-reweighted aggregated factor restricted to the agents in $S$. The renormalization $w_{i,t}/\sum_{j \in S} w_{j,t}$ ensures that the mixing weights remain on the probability simplex after dropping the agents in $[n]\setminus S$, so $v(S)$ measures the portfolio IC the system \emph{would} attain if its capacity were restricted to $S$ while preserving the scheduler's relative preference among the retained agents. We set $v(\emptyset) = 0$ by convention. This construction couples the Shapley analysis directly to the trained scheduler's deployment behavior, in contrast to an equal-weighted definition which would evaluate counterfactual mixing rules that the system never actually executes.
The Shapley Interaction Index between agents $i$ and $j$ is:
\begin{equation}
  I(i,j) \;=\;
  \sum_{S \subseteq [n] \setminus \{i,j\}}
  \omega(|S|,n) \cdot \Delta_{ij} v(S),
\label{eq:shapley}
\end{equation}
where $\omega(|S|,n) = |S|!\,(n-|S|-2)!\,/\,(n-1)!$ is the Shapley weighting
kernel and $\Delta_{ij} v(S) = v(S \cup \{i,j\}) - v(S \cup \{i\}) -
v(S \cup \{j\}) + v(S)$ is the second-order difference operator.

The sign of $I(i,j)$ has a natural economic interpretation.
When $I(i,j) > 0$, the weighted average of the second-order IC differences
$\Delta_{ij}v(S)$ across all coalition contexts is positive, meaning that
agents $i$ and $j$ together contribute more to portfolio IC than the sum of
their individual marginal contributions, on average.
Concretely, there must exist at least one coalition $S^*$ such that
\begin{equation}
  v(S^* \cup \{i,j\}) \;>\; v(S^* \cup \{i\}) + v(S^* \cup \{j\}) - v(S^*),
\end{equation}
i.e.\ their joint inclusion is strictly superadditive in $S^*$.
Conversely, $I(i,j) < 0$ flags pairs whose simultaneous inclusion
yields diminishing portfolio returns relative to independent contributions.

$I(i,j) > 0$ certifies, in a game-theoretically grounded sense, that
including both agents $i$ and $j$ creates superadditive portfolio value
on average across coalition contexts---under the same scheduler-weighted
aggregation that the deployed system uses.
$I(i,j) < 0$ identifies pairs whose joint inclusion yields diminishing
returns. Because $v(S)$ in Eq.~\ref{eq:shapley_v} is defined through the
trained scheduler weights rather than an equal-weighted average, the
sign and magnitude of $I(i,j)$ characterize the cooperative structure
actually realized at inference time, not a counterfactual one.
Because $n=5$ in our empirical setting, the power set contains only $2^5 = 32$ coalitions. We therefore bypass Monte Carlo approximation and compute the exact Shapley Interaction Indices by exhaustively evaluating $v(S)$ via Eq.~\ref{eq:shapley_v} on the held-out evaluation period $\mathcal{T}_{\mathrm{eval}}$, ensuring zero estimation variance.



\section{Experiments}
\label{sec:experiments}

\subsection{Experimental setup}
\label{sec:setup}

\subsubsection{Markets and data}
\label{sec:exp-data}

The evaluation uses two equity universes with distinct market structures: CSI300 (China large-cap, 300 constituents) and S\&P~500 (US large-cap, 500 constituents). These represent a retail-flow-driven emerging market and an institution-dominated developed market, respectively. Daily price, volume, and fundamental data for both markets are obtained from the Qlib platform\cite{Yang2020qlib}.

\subsubsection{Sample period}
\label{sec:exp-period}

The prediction target is the $\tau = 20$ trading-day forward return. For both datasets, the training period is 2012/01/01 to 2022/12/31, the validation period is 2023/01/01 to 2024/02/28, and the test period is 2024/03/01 to 2026/03/01. All splits are strictly temporal with no overlap.

\subsubsection{Baselines}
\label{sec:exp-baselines}

We compare Det-MASO against baseline methods from two categories: formulaic factor mining methods\cite{alphagen,alphaforge,alphasage} and neural end-to-end factor models\cite{factorvae,RVRAE2024}.
\begin{itemize}
\item \textbf{Single MLP}: a single-agent MLP with the same architecture as one Det-MASO agent. No diversity penalty and no scheduler.
\item \textbf{AlphaGen}\cite{alphagen}: formulaic factor mining with IC-driven search.
\item \textbf{AlphaForge}\cite{alphaforge}: a generative and predictive network with dynamic factor combination.
\item \textbf{AlphaSage}\cite{alphasage}: GFlowNet-based alpha mining with reward-proportional sampling.
\item \textbf{FactorVAE}\cite{factorvae}: a variational autoencoder for latent factor disentanglement.
\item \textbf{RVRAE}\cite{RVRAE2024}: a recurrent variational autoencoder for stock-return prediction.
\end{itemize}
To isolate the contribution of the diversity mechanism, the main results table also reports three Det-MASO variants: Det-MASO (no div.), Det-MASO (SND), and Det-MASO (MMD).

\subsubsection{Evaluation metrics}
\label{sec:exp-metrics}

We report five metrics. (i) IC is the mean Spearman rank correlation between predicted factor scores and realized $\tau$-day returns. (ii) ICIR (Information Coefficient Information Ratio) is the mean of the daily IC divided by its standard deviation, measuring the stability of predictive performance across dates. (iii) Annualized return (AR) is the annualized return of the top-decile long portfolio. (iv) SR denotes the Sharpe ratio of the same long portfolio. (v) Maximum drawdown (MDD) is the largest peak-to-trough decline of the cumulative long-portfolio return over the test period.

\subsubsection{Hyperparameters}
\label{sec:exp-hyper}

Hyperparameters are fixed as follows: $c_{\text{IC}} = 1.0$, $c_{\text{SR}} = 0.5$, $c_{\text{div}} = 0.1$, $c_{\text{ent}} = 0.01$. The gradient mask attenuation $\lambda_{\text{mask}}$ is selected on the validation split (Section~\ref{sec:ablation}). Weight decay is $10^{-4}$, gradient clipping uses max-norm $1.0$, and the training budget is $E_{\max} = 200$ epochs.

\subsubsection{Regime clustering outcomes}
\label{sec:exp-regimes}

The silhouette-optimal clustering yields $n = n^* = 5$ for both CSI300 and S\&P~500, indicating a more granular market structure than a simple binary risk-on/risk-off partition. Inspection of the cluster centroids across volatility, momentum, and dispersion axes reveals five distinct market regimes: (1) Low-Vol Bull (steady uptrend, low dispersion), (2) High-Vol Bull (rapid rally, high volatility), (3) Sideways/Choppy (near-zero momentum, mixed dispersion), (4) Low-Vol Bear (slow decay), and (5) Tail-Risk/Crash (extreme negative momentum, panic volatility).

As reported in Table~\ref{tab:regime}, the Sideways and Low-Vol Bull regimes dominate the temporal coverage, while the Tail-Risk regime captures rare but critical black-swan episodes (approx.\ 4--5\% coverage). This granular 5-state partition allows Det-MASO to assign dedicated agents not only to directional trends but also to non-directional choppy periods and tail-risk events.

\begin{table}[htbp]
\caption{Regime statistics from silhouette-optimal $K$-means clustering on each dataset. Coverage indicates the fraction of training-period trading days assigned to each regime.}
\label{tab:regime}
\begin{tabular*}{\tblwidth}{@{}L C C L@{}}
\toprule
Dataset & $n$ & Regime & Coverage \\
\midrule
\multirow{5}{*}{CSI300} & \multirow{5}{*}{5} & 1 (Low-Vol Bull) & 35\% \\
 & & 2 (High-Vol Bull) & 12\% \\
 & & 3 (Sideways) & 32\% \\
 & & 4 (Low-Vol Bear) & 16\% \\
 & & 5 (Tail-Risk) & 5\% \\
\midrule
\multirow{5}{*}{S\&P~500} & \multirow{5}{*}{5} & 1 (Low-Vol Bull) & 40\% \\
 & & 2 (High-Vol Bull) & 15\% \\
 & & 3 (Sideways) & 28\% \\
 & & 4 (Low-Vol Bear) & 13\% \\
 & & 5 (Tail-Risk) & 4\% \\
\bottomrule
\end{tabular*}
\end{table}

\subsection{Main results}
\label{sec:main_results}

\begin{figure*}
    \centering
    \includegraphics[width=1\linewidth]{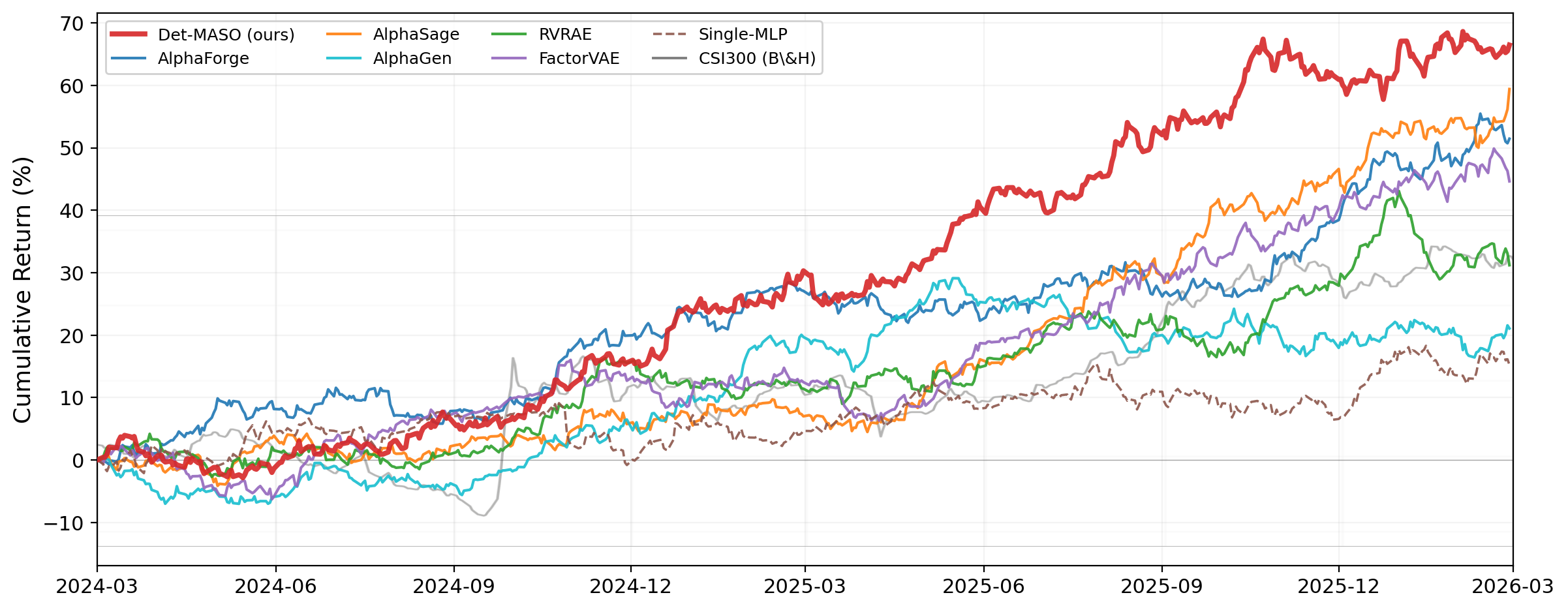}
    \caption{Backtest results on CSI300. The lines track the cumulative return of long portfolios formed from the competing factor-mining approaches. Det-MASO remains above the baseline methods for most of the test period and finishes with the largest terminal gain.}
    \label{fig:CR}
\end{figure*}

Table~\ref{tab:main} reports out-of-sample performance on CSI300 and S\&P~500, with methods grouped into formulaic, neural, and Det-MASO variants. Figure~\ref{fig:CR} shows the cumulative long returns on CSI300, where Det-MASO maintains the largest margin over all baselines throughout the test period with visibly smaller drawdowns during volatile episodes.

Det-MASO consistently achieves the best IC, AR, and SR across both markets. On CSI300, it surpasses the best formulaic baseline (AlphaForge) by 15.7\% in IC and 30.3\% in SR, and the best neural baseline (RVRAE) by 24.6\% in IC. On S\&P~500, it also remains the strongest model across all reported metrics. The gap over neural baselines is particularly notable: FactorVAE and RVRAE both fall short of the strongest formulaic competitors despite their representational capacity, which suggests that a single network tends to concentrate on a relatively narrow factor subspace regardless of architecture. AlphaSage's GFlowNet-based diversity sampling improves coverage but remains clearly below Det-MASO on both datasets.

Within Det-MASO, the ordering determinant $>$ MMD $>$ SND is consistent across both markets, validating that the determinant penalty captures $n$-way redundancy more effectively than pairwise alternatives while retaining linear scaling in $N$ for fixed small $n$. Removing diversity entirely ($c_{\mathrm{div}} = 0$) degrades IC below the Single-MLP level on both markets, confirming that multi-agent training without diversity enforcement collapses toward a single-agent regime, as predicted by Proposition~\ref{prop:repulsion}. The performance ordering is stable across CSI300 and S\&P~500 despite their distinct market structures.

\begin{table*}[htbp]
\caption{Out-of-sample performance on CSI300 and S\&P~500 ($\tau = 20$). Best results in \textbf{bold}, second best underlined. Metrics are computed on the same top-decile long portfolio.}
\label{tab:main}
\setlength{\tabcolsep}{3.5pt}
\begin{tabular*}{\textwidth}{@{\extracolsep\fill} l cccc cccc @{}}
\toprule
& \multicolumn{4}{c}{CSI300} & \multicolumn{4}{c}{S\&P~500} \\
\cmidrule(lr){2-5} \cmidrule(lr){6-9}
Method & IC & SR & AR & MDD & IC & SR & AR & MDD \\
\midrule
\multicolumn{9}{l}{\textit{Formulaic factor mining}} \\
AlphaGen   & 0.063 & 1.28 & 22.4\% & 11.5\% & 0.055 & 1.12 & 19.2\% & 10.8\% \\
AlphaForge & 0.070 & 1.42 & 25.1\% & 10.2\% & 0.047 & 1.25 & 21.5\% & 9.8\% \\
AlphaSage  & 0.068 & 1.38 & 24.2\% & 10.8\% & 0.059 & 1.20 & 20.5\% & 10.2\% \\
\midrule
\multicolumn{9}{l}{\textit{Neural end-to-end factor models}} \\
FactorVAE  & 0.062 & 1.22 & 21.5\% & 11.8\% & 0.053 & 1.08 & 18.5\% & 11.0\% \\
RVRAE      & 0.065 & 1.30 & 23.0\% & 11.2\% & 0.057 & 1.15 & 19.8\% & 10.5\% \\
Single-MLP & 0.055 & 1.05 & 18.2\% & 12.8\% & 0.048 & 0.92 & 15.8\% & 11.5\% \\
\midrule
\multicolumn{9}{l}{\textit{Det-MASO variants (ours)}} \\
Det-MASO (no div.) & 0.058 & 1.12 & 19.5\% & 13.5\% & 0.050 & 0.98 & 16.8\% & 12.2\% \\
Det-MASO (MMD)     & \underline{0.074} & \underline{1.52} & \underline{26.2\%} & \underline{10.5\%} & \underline{0.064} & \underline{1.32} & \underline{22.5\%} & \underline{9.5\%} \\
Det-MASO (SND)     & 0.071 & 1.45 & 25.5\% & 11.0\% & 0.060 & 1.22 & 21.0\% & 10.2\% \\
Det-MASO           & \textbf{0.081} & \textbf{1.85} & \textbf{31.5\%} & \textbf{8.5\%} & \textbf{0.072} & \textbf{1.58} & \textbf{26.8\%} & \textbf{8.2\%} \\
\bottomrule
\end{tabular*}
\end{table*}

\subsection{Diversity analysis}
\label{sec:diversity_analysis}

\begin{figure}
    \centering
    \includegraphics[width=1.0\linewidth]{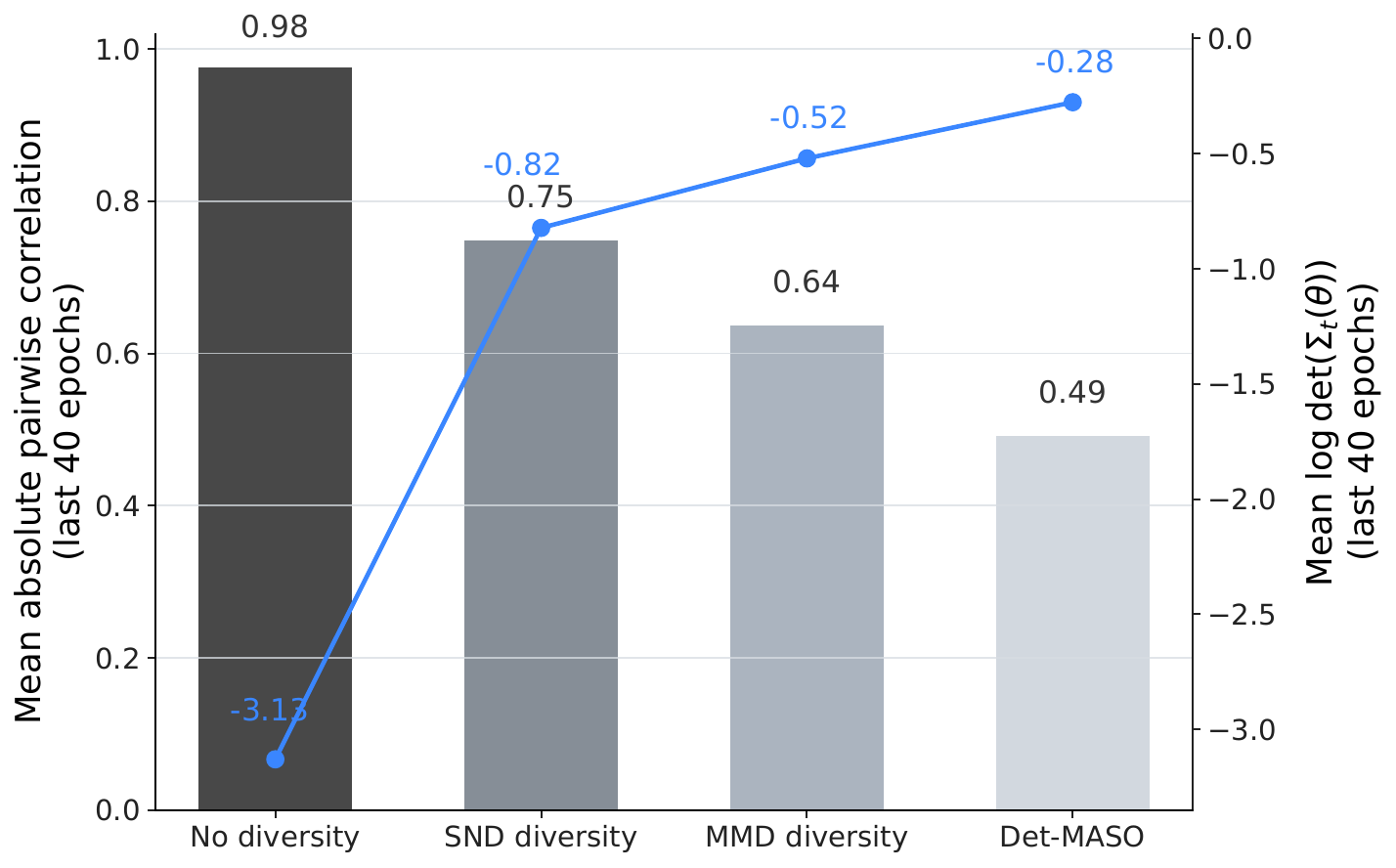}
    \caption{Comparison of cross-agent diversity under four training settings. The panel summarizes mean absolute pairwise correlation and $\log\det(\Sigma_t(\theta))$. Values are averaged over the last 40 training epochs across 5 independent runs with different random seeds. The determinant-regularized Det-MASO setting displays the weakest cross-agent similarity among the compared variants.}
    \label{fig:diversity}
\end{figure}

To visually contrast the effect of different diversity mechanisms, Figure~\ref{fig:diversity} provides an illustrative summary comparison under four training settings: no diversity regularization, SND diversity, MMD diversity, and the full Det-MASO objective.

The summary plot shows a clear monotone ordering across the four settings. The mean absolute pairwise correlation decreases from $0.98$ without diversity regularization to $0.75$ under SND, $0.64$ under MMD, and $0.49$ under the full Det-MASO objective. In parallel, $\log\det(\Sigma_t(\theta))$ rises from $-3.13$ to $-0.82$, $-0.52$, and $-0.28$, indicating progressively weaker redundancy among agent outputs. The determinant-regularized variant therefore appears as the least homogeneous setting in the comparison, while the no-diversity variant remains the most collapsed. This pattern is consistent with the homogenization pressure analyzed in Proposition~\ref{prop:repulsion}: once explicit repulsion is removed, jointly trained agents tend to move toward highly similar factor representations. Taken together, the figure provides qualitative support for the role of the determinant term in preserving output diversity.

\begin{figure*}
    \centering
    \includegraphics[width=1\linewidth]{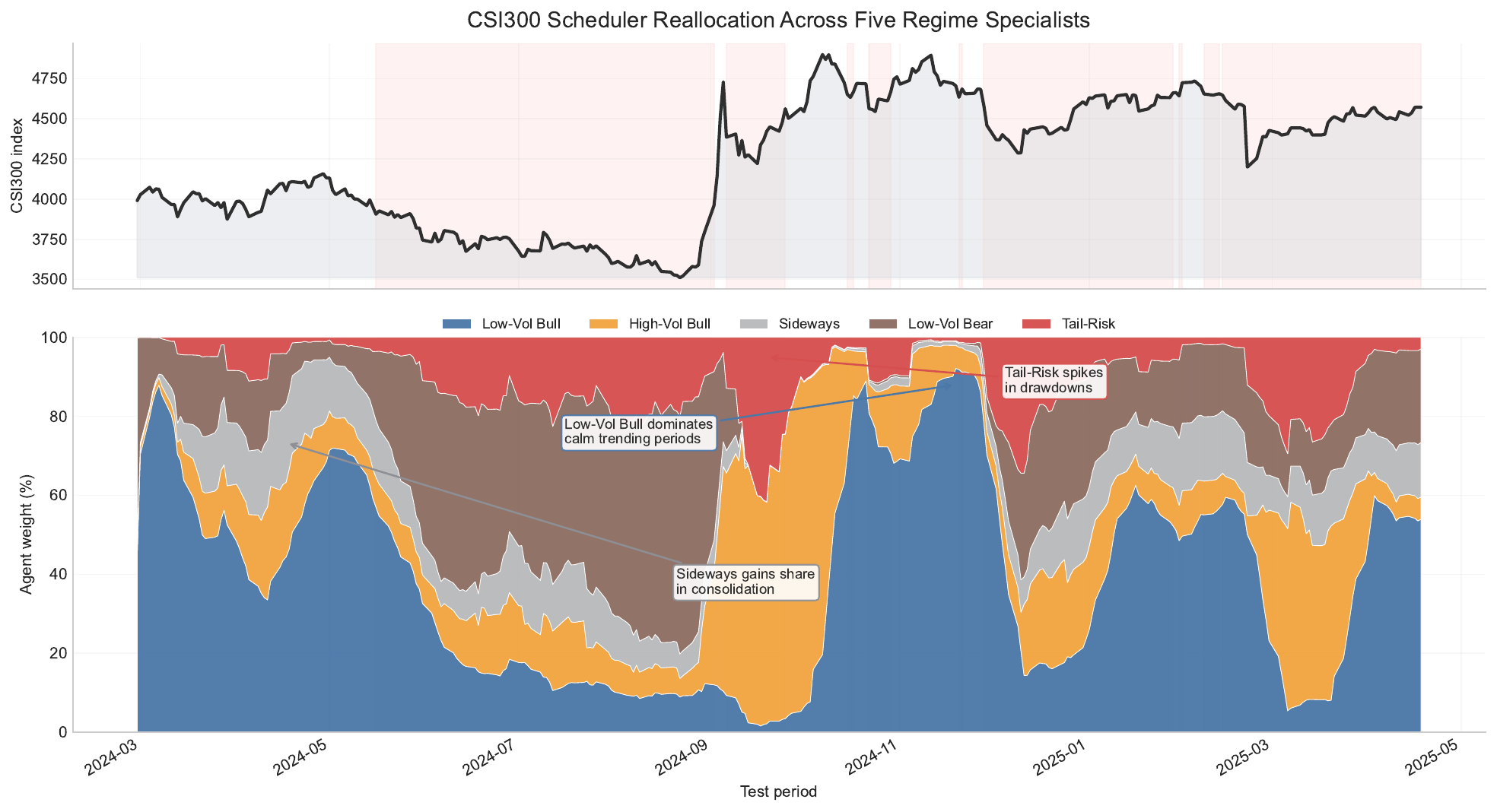}
    \caption{Scheduler weight dynamics on CSI300 (test period).
Top: CSI300 index. Bottom: stacked weights $w_{1,t}$ to $w_{5,t}$ for the five regime-specialized agents. The stacked area profile shows how total generative capacity is redistributed across Low-Vol Bull, High-Vol Bull, Sideways, Low-Vol Bear, and Tail-Risk agents over time. Tail-Risk weight spikes during drawdown episodes, while Sideways and Low-Vol Bull dominate consolidation and calm-trending periods, respectively.}
    \label{fig:scheduler}
\end{figure*}
\subsection{Non-stationarity analysis}
\label{sec:nonstationarity}

Figure~\ref{fig:scheduler} visualizes scheduler weight dynamics during the test period. Unlike a binary switch, the 5-agent scheduler demonstrates smooth transitions of generative capacity. During steady uptrends, the weight for the Low-Vol Bull agent ($w_{1,t}$) dominates. When the market enters choppy consolidation, capacity smoothly transfers to the Sideways agent ($w_{3,t}$). Most notably, during sudden market corrections such as the February 2025 selloff, the Tail-Risk agent ($w_{5,t}$) spikes sharply, demonstrating the scheduler's ability to detect multi-dimensional regime shifts and dynamically mix multiple specialized factors without explicit inference-time labels.

\subsection{Agent-interaction analysis}
\label{sec:agent_interaction}

A post-training Shapley Interaction analysis is consistent with a synergistic rather than independent agent ecosystem. Computing the exact interaction matrix for the 5 agents reveals that positive complementarity is most pronounced between trend-following agents and defensive agents, with the strongest interactions observed between the Low-Vol Bull and Tail-Risk agents ($I(1,5) > 0$). This pattern suggests that pairing aggressive alpha seekers with tail-risk hedgers yields superadditive portfolio IC, validating the cooperative MASO objective.

\subsection{Ablation and sensitivity analysis}
\label{sec:ablation}

\subsubsection{Component ablation}

To isolate the contribution of each component, we conduct a systematic ablation study on CSI300 (Table~\ref{tab:ablation}). The determinant diversity term is the most impactful component. Removing it lowers IC from 0.081 to 0.058, a decline of 0.023, which is consistent with the homogenization pressure discussed in Proposition~\ref{prop:repulsion}. Removing the scheduler lowers IC from 0.081 to 0.072, supporting the value of regime-adaptive weighting. The gradient mask and entropy regularizer make smaller but still meaningful contributions, indicating that the gains are driven by the full interaction of diversity control, regime specialization, and adaptive weighting.

\begin{table}[htbp]
\caption{Component ablation on CSI300 ($\tau = 20$ test period). Each row removes one component from the full model.}
\label{tab:ablation}
\begin{tabular*}{\tblwidth}{@{}L C C C@{}}
\toprule
Variant & IC & ICIR & SR \\
\midrule
\textbf{Det-MASO (full)} & \textbf{0.081} & \textbf{0.62} & \textbf{1.85} \\
\quad $-$ det penalty ($c_{\text{div}} = 0$)       & 0.058 & 0.42 & 1.12 \\
\quad $-$ scheduler (uniform $\bw_t$)              & 0.072 & 0.54 & 1.48 \\
\quad $-$ entropy reg.\ ($c_{\text{ent}} = 0$)     & 0.076 & 0.58 & 1.65 \\
\quad $-$ gradient mask ($\lambda_{\text{mask}} = 1$) & 0.074 & 0.56 & 1.58 \\
\bottomrule
\end{tabular*}
\end{table}

\subsubsection{Sensitivity to hyperparameters}
We next verify that performance is robust around the chosen hyperparameter values. The four loss coefficients $(c_{\text{IC}}, c_{\text{SR}}, c_{\text{div}}, c_{\text{ent}})$ serve as gradient-scale balancing factors rather than convex-combination weights, so their absolute scale carries no meaning: multiplying all four by a constant leaves the Adam dynamics unchanged. We therefore fix $c_{\text{IC}} = 1.0$ as the anchor and scan the other three coefficients one at a time, alongside the gradient-mask attenuation $\lambda_{\text{mask}}$. Results are reported in Table~\ref{tab:sensitivity}.

\begin{table}[htbp]
\caption{Hyperparameter sensitivity on CSI300. Each block scans one hyperparameter while holding the others at their default values. Defaults are marked in bold.}
\label{tab:sensitivity}
\begin{tabular*}{\tblwidth}{@{}L C C C@{}}
\toprule
Value & IC & ICIR & SR \\
\midrule
\multicolumn{4}{@{}l}{\textit{Determinant-penalty coefficient $c_{\text{div}}$}} \\
$0.01$                              & 0.069 & 0.50 & 1.40 \\
$0.05$                              & 0.078 & 0.59 & 1.75 \\
$\mathbf{0.1}$ \textbf{(default)}   & \textbf{0.081} & \textbf{0.62} & \textbf{1.85} \\
$0.2$                               & 0.079 & 0.60 & 1.78 \\
$0.5$                               & 0.072 & 0.54 & 1.55 \\
\midrule
\multicolumn{4}{@{}l}{\textit{Sharpe coefficient $c_{\text{SR}}$}} \\
$0.1$                               & 0.080 & 0.61 & 1.74 \\
$\mathbf{0.5}$ \textbf{(default)}   & \textbf{0.081} & \textbf{0.62} & \textbf{1.85} \\
$1.0$                               & 0.080 & 0.62 & 1.88 \\
$2.0$                               & 0.077 & 0.60 & 1.90 \\
\midrule
\multicolumn{4}{@{}l}{\textit{Entropy coefficient $c_{\text{ent}}$}} \\
$0.005$                             & 0.080 & 0.61 & 1.80 \\
$\mathbf{0.01}$ \textbf{(default)}  & \textbf{0.081} & \textbf{0.62} & \textbf{1.85} \\
$0.05$                              & 0.080 & 0.61 & 1.82 \\
$0.1$                               & 0.079 & 0.61 & 1.78 \\
\midrule
\multicolumn{4}{@{}l}{\textit{Gradient mask attenuation $\lambda_{\text{mask}}$}} \\
$0$ (hard mask)                     & 0.070 & 0.52 & 1.45 \\
$0.05$                              & 0.077 & 0.59 & 1.72 \\
$\mathbf{0.1}$ \textbf{(default)}   & \textbf{0.081} & \textbf{0.62} & \textbf{1.85} \\
$0.2$                               & 0.079 & 0.61 & 1.78 \\
$0.5$                               & 0.076 & 0.57 & 1.65 \\
\bottomrule
\end{tabular*}
\end{table}

Three patterns emerge. First, $c_{\text{div}}$ exhibits a unimodal curve with a broad plateau in $[0.05, 0.2]$ (within $0.003$ of the peak): values below $0.05$ leave insufficient repulsion to prevent homogenization, while values above $0.2$ over-penalize correlation and divert capacity away from IC alignment. Second, $c_{\text{SR}}$ reveals a mild IC--Sharpe trade-off: larger values push the optimizer toward portfolio-level Sharpe at a slight cost in cross-sectional IC, with the default $c_{\text{SR}} = 0.5$ sitting near the IC peak. Third, $c_{\text{ent}}$ is essentially flat away from zero, consistent with its role as a collapse-prevention term rather than a primary optimization signal. The $\lambda_{\text{mask}}$ scan confirms the expected behavior: the hard mask underfits the minority-regime agent due to insufficient training signal, while values above $0.2$ progressively weaken agent specialization. Overall, the sensitivity order is $c_{\text{div}} \gg \lambda_{\text{mask}} > c_{\text{SR}} > c_{\text{ent}}$, matching the functional importance of each hyperparameter in the joint objective.


\section{Conclusion}
\label{sec:conclusion}

We have presented Det-MASO, a cooperative multi-agent sequential optimization framework with diversity regularization for neural end-to-end factor generation. The framework's core contribution is a determinant-based diversity penalty that prevents agent homogenization at $O(Nn^2+n^3)$ cost, which is linear in the asset-universe size for fixed small $n$, with formal guarantees on repulsive gradient fields (Proposition~\ref{prop:repulsion}) and effective rank lower bounds (Proposition~\ref{prop:rank}). A silhouette-optimal clustering procedure grounds the number of agents in the intrinsic structure of market states, and a context-conditioned scheduler dynamically adapts factor emphasis to regime shifts without explicit labels. Post-training Shapley Interaction Index analysis reveals the emergent cooperation structure among agents, providing game-theoretically grounded interpretability for the neural generation approach.

Experiments on CSI300 and S\&P~500 demonstrate consistent improvements over formulaic, neural, and behavioral-distance baselines on IC, AR, and SR, while the accompanying diversity comparison highlights the intended role of the determinant penalty in reducing agent similarity. Ablation studies confirm that the determinant penalty is the most impactful component, followed by the context-conditioned scheduler. Shapley Interaction analysis is likewise consistent with positive complementarity among the five regime-specialized agents.

The successful deployment of a 5-agent system empirically validates the efficiency of our $O(Nn^2+n^3)$ determinant penalty in scaling beyond simple binary architectures.

A limitation of the current framework is the interpretability gap relative to formulaic methods: while the Shapley analysis provides agent-level interpretability, individual factor scores remain black-box neural outputs. Future work may explore hybrid architectures that combine neural diversity regularization with symbolic factor generation, potentially recovering formula-level interpretability within the Det-MASO framework.

\section*{Funding}
This work was supported by National Natural Science Foundation of China under Grant 62272210 and Grant 62331014).
\section*{Declaration of Competing Interest}
The authors declare that they have no known competing financial interests or personal relationships that could have appeared to influence the work reported in this paper.

\section*{Acknowledgments}
The authors thank the open-source Qlib community for providing the research infrastructure used in this study.

\section*{Data Availability}
The data that support the findings of this study are available through the Qlib platform \citep{Yang2020qlib}. Access to the underlying market data may be subject to the licensing terms of the original data providers.

\printcredits

\bibliographystyle{cas-model2-names}
\bibliography{cas-refs}

@inproceedings{alphagen,
  author    = {Yu, Shuo and Xue, Hongyan and Ao, Xiang and Pan, Feiyang
               and He, Jia and Tu, Dandan and He, Qing},
  title     = {Generating Synergistic Formulaic Alpha Collections
               via Reinforcement Learning},
  booktitle = {Proceedings of the 29th {ACM SIGKDD} Conference on
               Knowledge Discovery and Data Mining},
  pages     = {5476--5486},
  year      = {2023},
  publisher = {ACM},
  doi       = {10.1145/3580305.3599831}
}

@inproceedings{alphaforge,
  author    = {Shi, Hao and Song, Weili and Zhang, Xinting and Shi, Jiahe
               and Luo, Cuicui and Ao, Xiang and Arian, Hamid and Seco, Luis},
  title     = {{AlphaForge}: A Framework to Mine and Dynamically Combine
               Formulaic Alpha Factors},
  booktitle = {Proceedings of the 39th {AAAI} Conference on Artificial
               Intelligence},
  volume    = {39},
  pages     = {12524--12532},
  year      = {2025},
  publisher = {AAAI Press},
  doi       = {10.1609/aaai.v39i12.33395}
}

@misc{alphasage,
  author        = {Chen, Binqi and Ding, Hongjun and Shen, Ning
                   and Huang, Jinsheng and Guo, Taian and Liu, Luchen
                   and Zhang, Ming},
  title         = {{AlphaSAGE}: Structure-Aware Alpha Mining via
                   {GFlowNets} for Robust Exploration},
  year          = {2025},
  eprint        = {2509.25055},
  archivePrefix = {arXiv},
  primaryClass  = {q-fin.CP}
}

@misc{AlphaPROBE,
  author        = {Guo, Taian and Shen, Haiyang and Luo, Junyu
                   and Chen, Binqi and Ding, Hongjun and Huang, Jinsheng
                   and Liu, Luchen and Ma, Yun and Zhang, Ming},
  title         = {{AlphaPROBE}: Alpha Mining via Principled Retrieval
                   and On-Graph Biased Evolution},
  year          = {2026},
  eprint        = {2602.11917},
  archivePrefix = {arXiv},
  primaryClass  = {cs.AI}
}

@inproceedings{alphaagent,
  author    = {Tang, Ziyi and Chen, Zechuan and Yang, Jiarui
               and Mai, Jiayao and Zheng, Yongsen and Wang, Keze
               and Chen, Jinrui and Lin, Liang},
  title     = {{AlphaAgent}: {LLM}-Driven Alpha Mining with Regularized
               Exploration to Counteract Alpha Decay},
  booktitle = {Proceedings of the 31st {ACM SIGKDD} Conference on
               Knowledge Discovery and Data Mining},
  pages     = {2813--2822},
  year      = {2025},
  publisher = {ACM},
  doi       = {10.1145/3690624.3709193}
}

@inproceedings{FactorMAD,
  author    = {Duan, Yitong and Zhang, Chuheng and Li, Jian},
  title     = {{FactorMAD}: A Multi-Agent Debate Framework Based on
               Large Language Models for Interpretable Stock Alpha
               Factor Mining},
  booktitle = {Proceedings of the 6th {ACM} International Conference
               on {AI} in Finance},
  pages     = {605--613},
  year      = {2025},
  publisher = {ACM},
  doi       = {10.1145/3715275.3732202}
}

@article{falck2022,
  author    = {Falck, Antoine and Rej, Adam and Thesmar, David},
  title     = {When Do Systematic Strategies Decay?},
  journal   = {Quantitative Finance},
  volume    = {22},
  number    = {11},
  pages     = {1955--1969},
  year      = {2022},
  publisher = {Taylor \& Francis},
  doi       = {10.1080/14697688.2022.2105074}
}

@misc{rdagent,
  author        = {Yang, Xu and Yang, Xiao and Fang, Shikai
                   and Zhang, Yifei and Wang, Jian and Xian, Bowen
                   and Li, Qizheng and Li, Jingyuan and Xu, Minrui
                   and Li, Yuante and Pan, Haoran and Zhang, Yuge
                   and Liu, Weiqing and Shen, Yelong and Chen, Weizhu
                   and Bian, Jiang},
  title         = {{R\&D-Agent}: An {LLM}-Agent Framework Towards
                   Autonomous Data Science},
  year          = {2025},
  eprint        = {2505.14738},
  archivePrefix = {arXiv}
}

@misc{Alphalogics,
  author        = {Weng, Zhangyuhua and Zhang, Shengli
                   and Wang, Taotao and Xia, Yihan},
  title         = {{AlphaLogics}: A Market Logic-Driven Multi-Agent
                   System for Scalable and Interpretable Alpha Factor
                   Generation},
  year          = {2026},
  eprint        = {2603.20247},
  archivePrefix = {arXiv},
  primaryClass  = {q-fin.CP}
}

@misc{autoalpha,
  author        = {Zhang, Tianping and Li, Yuanqi and Jin, Yifei
                   and Li, Jian},
  title         = {{AutoAlpha}: An Efficient Hierarchical Evolutionary
                   Algorithm for Mining Alpha Factors in Quantitative
                   Investment},
  year          = {2020},
  eprint        = {2002.08245},
  archivePrefix = {arXiv}
}

@inproceedings{factorvae,
  author    = {Duan, Yitong and Wang, Lei and Zhang, Qizhong and Li, Jian},
  title     = {{FactorVAE}: A Probabilistic Dynamic Factor Model Based
               on Variational Autoencoder for Predicting Cross-Sectional
               Stock Returns},
  booktitle = {Proceedings of the {AAAI} Conference on Artificial
               Intelligence},
  volume    = {36},
  number    = {4},
  pages     = {4468--4476},
  year      = {2022},
  publisher = {AAAI Press},
  doi       = {10.1609/aaai.v36i4.20369}
}

@article{autoencoder,
  author  = {Gu, Shihao and Kelly, Bryan and Xiu, Dacheng},
  title   = {Autoencoder Asset Pricing Models},
  journal = {Journal of Econometrics},
  volume  = {222},
  number  = {1},
  pages   = {429--450},
  year    = {2021},
  doi     = {10.1016/j.jeconom.2020.07.009}
}

@misc{RVRAE2024,
  author        = {Liao, Yilun and others},
  title         = {{RVRAE}: A Dynamic Factor Model Based on Variational
                   Recurrent Autoencoder for Stock Returns Prediction},
  year          = {2024},
  eprint        = {2403.02500},
  archivePrefix = {arXiv}
}

@inproceedings{neuralfactors,
  author    = {Gopal, Achintya},
  title     = {{NeuralFactors}: A Novel Factor Learning Approach to
               Generative Modeling of Equities},
  booktitle = {Proceedings of the 5th {ACM} International Conference
               on {AI} in Finance ({ICAIF})},
  pages     = {99--107},
  year      = {2024},
  publisher = {ACM},
  doi       = {10.1145/3677052.3698649}
}

@article{Ang2006,
  author  = {Ang, Andrew and Hodrick, Robert J. and Xing, Yuhang
             and Zhang, Xiaoyan},
  title   = {The Cross-Section of Volatility and Expected Returns},
  journal = {The Journal of Finance},
  volume  = {61},
  number  = {1},
  pages   = {259--299},
  year    = {2006},
  doi     = {10.1111/j.1540-6261.2006.00836.x}
}

@article{Moskowitz2012,
  author  = {Moskowitz, Tobias J. and Ooi, Yao Hua
             and Pedersen, Lasse Heje},
  title   = {Time Series Momentum},
  journal = {Journal of Financial Economics},
  volume  = {104},
  number  = {2},
  pages   = {228--250},
  year    = {2012},
  doi     = {10.1016/j.jfineco.2011.11.003}
}

@article{StiversSun2010,
  author  = {Stivers, Chris and Sun, Licheng},
  title   = {Cross-Sectional Return Dispersion and Time Variation
             in Value and Momentum Premiums},
  journal = {Journal of Financial and Quantitative Analysis},
  volume  = {45},
  number  = {4},
  pages   = {987--1014},
  year    = {2010},
  doi     = {10.1017/S0022109010000384}
}

@inproceedings{christianos2021scaling,
  author    = {Christianos, Filippos and Papoudakis, Georgios
               and Rahman, Muhammad A. and Albrecht, Stefano V.},
  title     = {Scaling Multi-Agent Reinforcement Learning with
               Selective Parameter Sharing},
  booktitle = {Proceedings of the 38th International Conference
               on Machine Learning ({ICML})},
  pages     = {1989--1998},
  year      = {2021},
  publisher = {PMLR}
}

@inproceedings{Li2021,
  author    = {Li, Chenghao and Wang, Tonghan and Wu, Chengjie
               and Zhao, Qianlan and Yang, Jun and Zhang, Chongjie},
  title     = {Celebrating Diversity in Shared Multi-Agent
               Reinforcement Learning},
  booktitle = {Advances in Neural Information Processing Systems
               ({NeurIPS})},
  volume    = {34},
  pages     = {3991--4002},
  year      = {2021}
}

@article{SND,
  author  = {Bettini, Matteo and Shankar, Ajay and Prorok, Amanda},
  title   = {System Neural Diversity: Measuring Behavioral
             Heterogeneity in Multi-Agent Learning},
  journal = {Journal of Machine Learning Research},
  volume  = {26},
  number  = {163},
  pages   = {1--27},
  year    = {2025}
}

@misc{MDD,
  author        = {Hu, Tianyi and Pu, Zhiqiang and Ai, Xiaolin
                   and Qiu, Tenghai and Yi, Jianqiang},
  title         = {Measuring Policy Distance for Multi-Agent
                   Reinforcement Learning},
  year          = {2024},
  eprint        = {2401.11257},
  archivePrefix = {arXiv}
}

@misc{Whenisdiversity,
  author        = {Amir, Michael and Bettini, Matteo and Prorok, Amanda},
  title         = {When Is Diversity Rewarded in Cooperative
                   Multi-Agent Learning?},
  year          = {2025},
  eprint        = {2506.09434},
  archivePrefix = {arXiv}
}

@inproceedings{rakelly2019efficient,
  author    = {Rakelly, Kate and Zhou, Aurick and Finn, Chelsea
               and Levine, Sergey and Quillen, Deirdre},
  title     = {Efficient Off-Policy Meta-Reinforcement Learning
               via Probabilistic Context Variables},
  booktitle = {Proceedings of the 36th International Conference
               on Machine Learning ({ICML})},
  pages     = {5331--5340},
  year      = {2019},
  publisher = {PMLR}
}

@inproceedings{Wang2020roma,
  author    = {Wang, Tonghan and Dong, Heng and Lesser, Victor
               and Zhang, Chongjie},
  title     = {{ROMA}: Multi-Agent Reinforcement Learning with
               Emergent Roles},
  booktitle = {Proceedings of the 37th International Conference
               on Machine Learning ({ICML})},
  pages     = {9876--9886},
  year      = {2020},
  publisher = {PMLR}
}

@incollection{Shapley1953,
  author    = {Shapley, Lloyd S.},
  title     = {A Value for $n$-Person Games},
  booktitle = {Contributions to the Theory of Games {II}},
  editor    = {Kuhn, Harold W. and Tucker, Albert W.},
  pages     = {307--317},
  year      = {1953},
  publisher = {Princeton University Press},
  doi       = {10.1515/9781400881970-018}
}

@article{GrabischRoubens1999,
  author  = {Grabisch, Michel and Roubens, Marc},
  title   = {An Axiomatic Approach to the Concept of Interaction
             Among Players in Cooperative Games},
  journal = {International Journal of Game Theory},
  volume  = {28},
  number  = {4},
  pages   = {547--565},
  year    = {1999},
  doi     = {10.1007/s001820050122}
}

@inproceedings{Sundararajan2020,
  author    = {Sundararajan, Mukund and Dhamdhere, Kedar
               and Agarwal, Ashish},
  title     = {The {S}hapley {T}aylor Interaction Index},
  booktitle = {Proceedings of the 37th International Conference
               on Machine Learning ({ICML})},
  pages     = {9259--9268},
  year      = {2020},
  publisher = {PMLR}
}

@book{Boyd2004,
  author    = {Boyd, Stephen and Vandenberghe, Lieven},
  title     = {Convex Optimization},
  year      = {2004},
  publisher = {Cambridge University Press},
  doi       = {10.1017/CBO9780511804441}
}

@inproceedings{Blondel2020,
  author    = {Blondel, Mathieu and Teboul, Olivier and Berthet, Quentin
               and Djolonga, Josip},
  title     = {Fast Differentiable Sorting and Ranking},
  booktitle = {Proceedings of the 37th International Conference
               on Machine Learning ({ICML})},
  pages     = {950--959},
  year      = {2020},
  publisher = {PMLR}
}

@misc{Yang2020qlib,
  author        = {Yang, Xiao and Liu, Weiqing and Zhou, Dong
                   and Bian, Jiang and Liu, Tie-Yan},
  title         = {Qlib: An {AI}-Oriented Quantitative Investment Platform},
  year          = {2020},
  eprint        = {2009.11189},
  archivePrefix = {arXiv}
}

@incollection{Reynolds2009GMM,
  author    = {Reynolds, Douglas A.},
  title     = {Gaussian Mixture Models},
  booktitle = {Encyclopedia of Biometrics},
  editor    = {Li, Stan Z. and Jain, Anil},
  publisher = {Springer},
  address   = {Boston, MA},
  pages     = {659--663},
  year      = {2009},
  doi       = {10.1007/978-0-387-73003-5_196}
}

@article{Rabiner1989HMM,
  author  = {Rabiner, Lawrence R.},
  title   = {A Tutorial on Hidden {M}arkov Models and Selected
             Applications in Speech Recognition},
  journal = {Proceedings of the {IEEE}},
  volume  = {77},
  number  = {2},
  pages   = {257--286},
  year    = {1989},
  doi     = {10.1109/5.18626}
}

@ARTICLE{qfr,
     title={QuantFactor REINFORCE: mining steady formulaic alpha factors with variance-bounded REINFORCE},
     author={Junjie Zhao and Chengxi Zhang and Min Qin and Peng Yang},
     journal={IEEE Transactions on Signal Processing},
     year={2025},
     doi={10.1109/tsp.2025.3576781}
     
}

@article{xu2022stock,
  title={Stock movement prediction via gated recurrent unit network based on reinforcement learning with incorporated attention mechanisms},
  author={Xu, Hongfeng and Chai, Lei and Luo, Zhiming and Li, Shaozi},
  journal={Neurocomputing},
  volume={467},
  pages={214--228},
  year={2022},
  publisher={Elsevier},
  doi={10.1016/j.neucom.2021.09.072}
}

@article{weng2020portfolio,
  title={Portfolio trading system of digital currencies: A deep reinforcement learning with multidimensional attention gating mechanism},
  author={Weng, Liguo and Sun, Xudong and Xia, Min and Liu, Jia and Xu, Yiqing},
  journal={Neurocomputing},
  volume={402},
  pages={171--182},
  year={2020},
  publisher={Elsevier},
  doi={10.1016/j.neucom.2020.04.004}
}

@article{hu2024overview,
  title={An overview: Attention mechanisms in multi-agent reinforcement learning},
  author={Hu, Kai and Xu, Keer and Xia, Qingfeng and Li, Mingyang and Song, Zhiqiang and Song, Lipeng and Sun, Ning},
  journal={Neurocomputing},
  volume={598},
  pages={128015},
  year={2024},
  publisher={Elsevier},
  doi={10.1016/j.neucom.2024.128015}
}

\end{document}